\documentclass[namedreferences,hyperref,optionalrh]{spr-sola}
\usepackage{graphicx}
\usepackage{gensymb}
\usepackage{color}           

\usepackage{lscape}
\usepackage{threeparttable}
\usepackage{textcomp, gensymb}
\usepackage{lmodern}

\chardef\us=`\_

\begin{document}

\begin{frontmatter}
\title{Citizen CATE 2024: Extending Totality During the 8 April 2024 Total Solar Eclipse with a Distributed Network of Community Participants}

\author[addressref={aff1,aff2},corref,email={sakovac@ucar.edu}]{\inits{S.A.}\fnm{Sarah~A.}~\snm{Kovac}\orcid{0000-0003-1714-5970}}
\author[addressref=aff1,email={amir.caspi@swri.org}]{\inits{A.}\fnm{Amir}~\snm{Caspi}\orcid{0000-0001-8702-8273}}
\author[addressref=aff1]{\inits{D.B.}\fnm{Daniel~B.}~\snm{Seaton}\orcid{0000-0002-0494-2025}}
\author[addressref=aff2]{\inits{P.}\fnm{Paul}~\snm{Bryans}\orcid{0000-0001-5681-9689}}
\author[addressref=aff2]{\inits{J.R.}\fnm{Joan~R.}~\snm{Burkepile}\orcid{0000-0002-9959-6048}}
\author[addressref=aff75]{\inits{S.}\fnm{Sarah~J.}~\snm{Davis}\orcid{0009-0008-4901-0601}}
\author[addressref=aff1]{\inits{C.E.}\fnm{Craig~E.}~\snm{DeForest}\orcid{0000-0002-7164-2786}}
\author[addressref=aff4]{\inits{D.}\fnm{David}~\snm{Elmore}\orcid{0009-0000-4413-6434}}
\author[addressref=aff4]{\inits{S.}\fnm{Sanjay}~\snm{Gosain}\orcid{0000-0002-5504-6773}}
\author[addressref=aff2]{\inits{R.}\fnm{Rebecca}~\snm{Haacker}\orcid{0000-0002-6501-8266}}
% ...

\address[id=aff1]{Southwest Research Institute, Boulder, Colorado, USA}
\address[id=aff2]{U.S. National Science Foundation National Center for Atmospheric Research, Boulder, Colorado, USA}
\address[id=aff75]{Montana State University, Bozeman, Montana}
\address[id=aff4]{National Solar Observatory, Boulder, Colorado, USA}
% ...

\runningauthor{S.A. Kovac et al.}
\runningtitle{Citizen CATE 2024}

\begin{abstract}
The Citizen CATE 2024 next-generation experiment placed 43 identical telescope and camera setups along the path of totality during the total solar eclipse (TSE) on 8~April~2024 to capture a 60-minute movie of the inner and middle solar corona in polarized visible light. The 2024~TSE path covered a large geographic swath of North America and we recruited and trained 36~teams of community participants (``citizen scientists'') representative of the various communities along the path of totality. Afterwards, these teams retained the equipment in their communities for on-going education and public engagement activities. Participants ranged from students (K12, undergraduate, and graduate), educators, and adult learners to amateur and professional astronomers. In addition to equipment for their communities, CATE~2024 teams received hands-on telescope training, educational and learning materials, and instruction on data analysis techniques. CATE~2024 used high-cadence, high-dynamic-range (HDR) polarimetric observations of the solar corona to characterize the physical processes that shape its heating, structure, and evolution at scales and sensitivities that cannot be studied outside of a TSE. Conventional eclipse observations do not span sufficient time to capture changing coronal topology, but the extended observation from CATE~2024 does. Analysis of the fully calibrated dataset will provide deeper insight and understanding into these critical physical processes. We present an overview of the CATE~2024 project, including how we engaged local communities along the path of totality, and the first look at CATE~2024 data products from the 2024~TSE.
\end{abstract}
\end{frontmatter}

\section{Introduction}\label{sec:intro} 
On 8~April~2024, the United States of America experienced its second total solar eclipse (TSE) in seven years. Over North America, the 2024 path of totality extended from Mexico into Texas, northeast through the Midwest and New England, and ended in the Canadian Maritimes. The entirety of the Moon's shadow's trip over the continental USA provided a view of the lower and middle corona for nearly 60~minutes.

The middle corona is a region of significant physical transitions, including changes in magnetic topology and plasma properties, between 1.5 and 6~R$_\odot$ \citep{West2023}. The 1.5--3~R$_\odot$ region, in particular, is dominated by complex dynamics where the magnetic field transitions from mostly closed loops to mostly open, radial structures in the vicinity of the streamer cusps, strongly modulating outflow into the solar wind \citep{Seaton2021, Chitta2023NatAs}. This region sets the connectivity between surface magnetic structures and the solar wind, but the transition between these two magnetic morphology regimes is not well understood, as the required resolution is too fine for existing space-based instruments to trace detailed connectivity through the cusp region. Observations during a total solar eclipse -- particularly polarimetric observations that can isolate K-corona structures in which the middle corona transitions occur -- are of prime value to understanding this sparsely studied region.

The innermost parts of the solar corona are notoriously difficult to observe in visible light. While X-ray and ultraviolet observations routinely reveal the inner and, recently, middle corona \citep{Seaton2021}, in visible light, the inner and middle corona can only be seen when there is an occulter blocking out the much brighter light coming from the photosphere. Coronagraphic observations of this region are challenging because, close to the solar disk, it is difficult to suppress stray light from the photosphere. Several instruments, including the Mauna Loa Solar Observatory K-Cor coronagraph \citep{dewijn_2012} and its many predecessors, and more recently, the \textit{Proba-3} formation-flying coronagraph \citep{Shestov2021} have made such observations \citep{zhukov2025}. However, natural total solar eclipses best provide the image contrast required to measure the finest structure of this region and the temporal evolution that can reveal answers to key questions about how the corona is heated and how the solar wind is accelerated. 

Unpolarized visible-light imaging has previously been used to infer the topology of the coronal magnetic field \citep{Boe2020} and even to measure the propagation of small coronal mass ejections (CMEs) \citep{Pasachoff2011}. Visible emission line observations have been used to characterize key parameters of coronal plasma \citep{Boe2018} and the processes that may heat it to millions of degrees \citep{Pasachoff2002}. However, because the polarization of the K-corona depends on the three-dimensional geometry of coronal plasma, polarimetric observations have been known to be an important tool for developing an understanding of the corona's 3D~nature for more than a century \citep{Young1911}. \citet{Billings_1966}, in particular, showed how the 3D~relationships between coronal structure and scattering angle can be encoded in the degree of polarization of visible light from the corona, and how the K-corona can be, to some extent, isolated from other, less polarized contributions like the F-corona via polarimetric analysis.

Polarimetric measurements of the inner and middle corona are even more challenging than white-light observations, due to the need for high signal-to-noise ratio at multiple polarization angles required to resolve the polarized brightness of the corona \citep{DeForest2022}. The newly-launched NASA \textit{Polarimeter to Unify the Corona and Heliosphere} (\textit{PUNCH}) mission uses a constellation of four microsatellites to produce 3D~images from 6~R$_\odot$ to 180~R$_\odot$ in polarized visible light \citep{deforest2025_punch}. However, the unique conditions during an eclipse are especially well-suited for polarimetric observations of the inner and middle corona (below 6~R$_\odot$).  Eclipses have been used for decades to probe the corona's 3D~structure \citep[e.g.,][]{Koutchmy1971}. These studies have sometimes produced contradictory results, indicating that the degree of polarization agrees with theoretical predictions for Thomson scattering \citep[e.g.,][]{Vorobiev2020} but also that it can exceed the theoretical maximum in some cases \citep[e.g.,][though such anomalous observations may be related to the polarization of light from the sky]{Merzlyakov2019}. Other observations from eclipses have facilitated the characterization of electron density across the corona and provided clues to the relationship between variations in the degree of polarization and 3D~structure of the corona as a function of height \citep{Liang2023}. 

Coronal evolution, including transient changes due to the propagation of waves within coronal structures \citep[e.g.,][]{Pasachoff2002} or magnetic reconnection associated with nanoflares and picojets \citep{Chitta2023, Raouafi2023}, can occur on various timescales, as short as seconds, to hours or more. In particular, morphological changes occur over longer time periods than a few minutes. However, because eclipses are transient phenomena, and most polarimetric observations have been made from only a single location, there have been very few studies that examined the evolution of polarization in the corona at small spatial and temporal scales. Extending the duration of polarized observations during an eclipse provides a means to characterize this evolution at the longer timescales required to understand its relationship to important physical processes in the corona. 

Many scientific expeditions have tried different methods to extend observing time, such as by trying to chase the shadow in a mobile observatory. An excellent example of this occurred during the 1973~TSE over Africa. The maximum eclipse duration at a given point on the ground would have allowed roughly seven minutes of totality observations, but by flying the Concorde at supersonic speed and following the shadow, the team collected 74~minutes of visible and IR observations of the corona \citep{beckman1973, Koutchmy1975}. More recently, multiple aircraft flying at different points along the path of totality have been used to extend totality observations. In 2017, \citet{caspi2020} flew two NASA WB-57 jets chasing the shadow of totality in tandem to lengthen the observing time to $\sim$7.6~minutes of coronal images in the visible and midwave IR (MWIR:~2--5~$\mu$m) spectrum, $\sim$3~times longer than a stationary ground-based observer would see. 

An alternative approach to extending totality observations is by distributing a network of ground-based observatories along the eclipse path. While coordinated, multi-location observations have been collected in the past \citep[e.g.,][]{pasachoff2009b, Penn2017}, both the 2017 and 2024 eclipses over North America provided rare opportunities in that the paths of totality crossed a densely populated and easily traversed region for an extended period of time. Due to the number of sites needed to maximize observations along the path and their geographically widespread locations, it is much more feasible for professional astronomers to do this work with the help of ``citizen scientists'' (hereafter, community participants). This opportunity was first leveraged by the 2017 Citizen CATE Experiment, an expedition which put 68~identical telescope setups teamed by community participants along the path of totality to capture a 90-minute movie of the solar corona in white light \citep{Penn2020}. A plethora of additional experiments with community participants were conducted in 2017, one example being Eclipse Mega\-Movie~2017, which aimed to create a 90-minute movie of totality by having participants across the path capture images via smartphones \citep{megamovie2022}. 

The successes of the 2017 distributed observing endeavors inspired even more participatory science projects in 2024 to extend totality observations. The leaders of the original 2017 Citizen CATE Experiment pioneered the Dynamic Eclipse Broadcast (DEB) Initiative, which was a ground-based, low-cost telescope network used to live-stream totality \citep{baer2022, mandrell2023, mandrell2024} and further advance the scientific goals described in \citet{Penn2020}. The Eclipse MegaMovie project ran again in 2024, but this time using DSLR cameras \citep{megamovie2024} instead of relying on cell phone images. Many new projects were developed, such as Sun\-Sketcher, which worked similarly to Eclipse Mega\-Movie~2017, where participants were asked to take images with their cell phone cameras, but with a scientific focus on second and third contact to help better characterize Baily's beads \citep{sunsketchers2024}. All of these projects, plus countless others, will contribute to a better understanding of the Sun's dynamics over longer timescales. Many other participatory science projects across multiple disciplines further highlight the importance of various observations during eclipses \citep[e.g.,][]{colligan2020, soundscapes2024}. 

Development of these distributed networks of community participants not only provides unique scientific opportunities, but also unites the communities who experience this celestial event together \citep{Pasachoff2009}, making these networks an excellent way to engage with local communities and individuals along the path, as demonstrated by the original 2017 Citizen CATE Experiment. In 2024, with a new project team, the next-generation CATE~2024 experiment followed in the footsteps of the 2017 Citizen CATE Experiment and used a large, well coordinated network of community participant teams distributed along the path of totality on 8~April~2024 to capture near-continuous images of totality for $\sim$60~minutes, this time in polarized light to address new scientific questions. 

This paper describes the overarching project goals (Section~\ref{sec:project_goals}), and project implementation, including how the CATE~2024 community was built (Section~\ref{sec:pi_team}), what instrumentation and procedures were used (Sections~\ref{sec:pi_instrument} \& \ref{sec:procedure}), and how community participants were trained (Section~\ref{sec:training}) leading up to the eclipse. We present a summary of 8~Apr~2024 (Section~\ref{sec:eclipse_day}), including a first look at CATE~2024 data (Section~\ref{sec:results}) and calibration techniques. We conclude with a summary of the project and future opportunities for CATE~2024 participants in Section~\ref{sec:discussion}. 

\section{Project Goals and Objectives}\label{sec:project_goals}
\subsection{Science Goals}\label{sec:pg_science}
The objectives below are uniquely suited to the high-resolution, high-sensitivity coronal imaging possible during a total solar eclipse. They also require longer-duration observations than are possible from a single, fixed eclipse viewing location, making them highly suitable for the CATE~2024 observations. Table~\ref{table:performance_requirements} describes the driving requirements and the actual performance relative to each scientific objective. Descriptions of the instrumentation and observing procedures used to meet these requirements are detailed in Section~\ref{sec:pi_instrument} and Section~\ref{sec:procedure}, respectively.

The brightness of the corona drops quickly with distance from the Sun \citep{golub_pasachoff2010}, so in order to see a more expanded view of the corona in a single image, various length exposures are needed. At shorter exposures, the outer corona is faint and blends in with the noise in the image, and at longer exposures, the innermost corona becomes saturated. This drives the need to create high dynamic range (HDR) images by combining multiple images of varying exposure times. CATE~2024 collected eight distinct, logarithmically-spaced exposure times (0.13, 0.40, 1.3, 4.0, 13, 40, 130, and 400~ms) to capture the full range of coronal brightness out to 3~R$_\odot$. 

\begin{table}[!ht]
\caption{Driving requirements and actual performance for CATE~2024 instrumentation and operating procedure selection. The required dynamic range for the overall experiment is determined by the steep falloff in coronal brightness across the required FOV \citep{golub_pasachoff2010}.}
\label{table:performance_requirements}
\addtolength{\tabcolsep}{-.1em}
\begin{tabular}{llll}

\hline
 & Driving Req. & Actual Performance & Science Goals \\
\hline
Spatial Resolution & $5''$ FWHM & $2.9''$ FWHM ($1.43''$/pix) & Obj. 1, 2, 3 \\
Field of View & $\pm$3\,R$_\odot$ (eq.) & $\pm$3.04 (eq.) $\times\,\pm$2.23 (pol.)\,R$_\odot$ & Obj. 1, 2, 3 \\
Dynamic Range & $10^{-10} - 10^{-5}$\,B$_\odot$  & $10^{-11}->$10$^{-4}$\,B$_\odot$ (HDR) & Obj. 1, 2, 3 \\
Polarization & 2--4\% & 1\% & Obj. 1, 2 \\
Sensitivity \\
Cadence & 1~minute & 1.6~s per HDR sequence & Obj. 2, 3 \\
Duration & 1~hour & 1h28m07s$^1$ & Obj. 1, 2, 3 \\
\end{tabular}
\begin{tablenotes}
   \item[*]$^1${This is the maximum duration between the start of totality at the first observing \\ site in Mexico and the end of totality at the last observing site in Maine.}
  \end{tablenotes}
\end{table}

\subsubsection{Objective 1: Determine connectivity of the middle corona}
A key feature of the middle corona that can be characterized using eclipse observations is the transition from mostly closed loops to mostly open, radial structures \citep{West2023, Seaton2023}. There is some evidence that, particularly in the vicinity of the streamer cusps,  interactions within this region modulate outflow into the solar wind \citep{Chitta2023NatAs}. However, the connectivity between surface magnetic structures and the solar wind, across the transition between open and closed magnetic morphology regimes, is not well understood. This objective seeks to use CATE~2024's high-cadence, high-resolution, polarimetric observations to characterize the topological transition across the corona, from open to closed, and examine the interface between these regions for flows or evolution that may be linked to solar wind outflow. These observations can be analyzed to infer 3D~structure and to characterize fine structure observed by the 2017 Citizen CATE Experiment \citep{Penn2020} and other projects \citep[e.g.,][]{deforest2013,deforest2017,DeForest2022}, along with direct measurements of electron density \citep{Sivaraman1984,Habbal2011}.

\subsubsection{Objective 2: Measure the nascent solar wind flow}
The association of slow solar wind flow with particular features in the low corona is a long-outstanding open question and distinguishes between major classes of currently-open solar wind models \citep[e.g.,][]{abbo2016, kepko2016}. The solar wind is highly inhomogeneous and can be tracked visually throughout the corona over time \citep[e.g.,][]{deforest2018}. Solar wind speeds at 2~R$_\odot$ are typically $\sim$50--150~km/s \citep[e.g.,][]{deforest1997,kohl1997}. Eclipse measurements at one fixed location last only a few minutes, making them essentially still frames with such a short interval compared to typical coronal evolution time scales. CATE~2024 captured nearly 60~consecutive minutes of observations, a sufficient duration to distinguish where the solar wind is flowing in the lower and middle corona. CATE~2024 will thereby be able to distinguish between models that include release or diffusion through the closed-field region, from those requiring release at coronal hole edges. 

\subsubsection{Objective 3: Identify and characterize reconnection in the middle corona}
Magnetic reconnection in the low to middle corona is essential to maintaining the smooth ``combed'' appearance of the corona as a whole and to nearly all coronal heating and wind origination models \citep[e.g.,][]{longcope2015}. This inferred non-flare reconnection in the ``bulk quiet corona'' has historically been invisible to most instruments. However, it has been detected with EUV imaging of bright, well-presented loops \citep[e.g.,][]{li2018} and is associated with changing morphology over minutes to hours. EUV observations of in-progress non-flare reconnection in the quiet corona are rare partly because of the low signal-to-noise ratios of existing EUV images and partly because of the narrow temperature sensitivity of collisionally excited (or resonantly scattered) emission lines. Thomson-scattered light offers a broad thermal response and direct sensitivity to line-of-sight density; this is reflected in the rich loop detail available in even moderately enhanced images \citep[e.g.,][]{srivastava2019}. Conventional eclipse observations do not span sufficient time to capture changing coronal topology, but the extended observation with CATE~2024 does, on timescales of a few minutes to an hour. 

\subsection{Education and Engagement}\label{sec:pg_education}
Broad-scale and authentic public education and community engagement was one of the primary objectives for CATE~2024. Since the path of totality covered a wide swath of the South Central, Midwest, and Northeast United States, including a variety of rural, urban, and Tribal communities, we focused on recruiting observing groups reflective of these communities. We specifically prioritized selecting teams who may not have access to scientific equipment or STEM resources. Additionally, we strove to select teams without prior astronomy experience. In this way, CATE~2024 served as an opportunity for local teams to see themselves as scientists, explore potential careers in STEM, and contribute to world-class scientific research.

To promote community-wide usage of the equipment, the local observing teams worked with the core team to identify locations where we could donate the CATE~2024 telescopes after the eclipse. This included locations such as museums, public libraries, community centers, K12 schools, and universities, so that they could become free perpetual community resources. To extend the utility of the telescopes, the CATE~2024 core team also developed a handbook of learning activities for both daytime and nighttime observing. 

\section{Project Implementation}\label{sec:project_implementation}
Distributed eclipse observations take extensive planning and coordination, particularly when a majority of the teams are led by amateur participants. Major project milestones and progression are described in Table~\ref{table:milestones}.

\begin{table}
\caption{Major project milestones. RC~=~Regional Coordinator. SC~=~State Coordinator. LT~=~Lead Trainer.}
\label{table:milestones}
\begin{tabular}{lll}
\hline
Date & Project Milestone & Location\\
\hline
Summer 2021--2022 & Preliminary Work & Boulder, CO \\
Fall 2022 & Recruit RCs \& LTs & Boulder, CO\\
1--2 Apr 2023 & Regional Training Workshop & Boulder, CO \\
11--22 Apr 2023 & TSE 2023 Expedition & Exmouth, WA, AUS \\
Fall 2023 & Recruit SCs \& LTs & Boulder, CO\\
14 Oct 2023 & ASE 2023 Observations & Albuquerque, NM \\
& & Loveland Pass, CO\\
22 Nov 2023 & Open application for local teams & Boulder, CO\\
19--21 Jan 2024 & Coordinator Training Workshop & San Antonio, TX \\
26--27 Jan 2024 & S. TX Local Team Workshop & Uvalde, TX \\
3--4 Feb 2024 & IN Local Team Workshop & Indianapolis, IN \\
& N. TX Local Team Workshop & Dallas, TX \\
10--11 Feb 2024 & MO/IL Local Team Workshop & Cape Girardeau, MO \\
& OK/AR Local Team Workshop & Conway, AR \\
16--17 Feb 2024 & NY Local Team Workshop & Rochester, NY \\
17 Feb 2024 & OH Local Team Workshop & Uniontown, OH \\
17--18 Feb 2024 & VT/NH/ME Local Team Workshop$^*$ & Orono, ME \\
24 Feb 2024 & First Practice Session & Nationwide \\
24--25 Feb 2024 & VT/NH/ME Local Team Workshop$^*$ & North Conway, NH \\
9 Mar 2024 & Second Practice Session & Nationwide \\
23 Mar 2024 & Third Practice Session & Nationwide \\
6 Apr 2024 & Optional Practice Opportunity & Nationwide \\
7 Apr 2024 & Final Practice Session & See Table~\ref{table:sites} \\
8 Apr 2024 & TSE 2024 Observations & See Table~\ref{table:sites}\\
Summer 2024 & Datasets Mailed to Science Team & Boulder, CO \\
& Preliminary Data Processing & Boulder, CO \\
10 Dec 2024 & Public Release of Preliminary Movie & Washington, DC \\ 
\hline
\end{tabular}
\begin{tablenotes}
   \item[*]$^*$Due to the large geographical separation between teams, it was in the \\
   participants' best interests to host separate workshops. 
  \end{tablenotes}
\end{table}

\subsection{Building the CATE~2024 Community}\label{sec:pi_team}
\begin{figure}[H]
    \centering
    \centerline{\includegraphics[width=\textwidth]{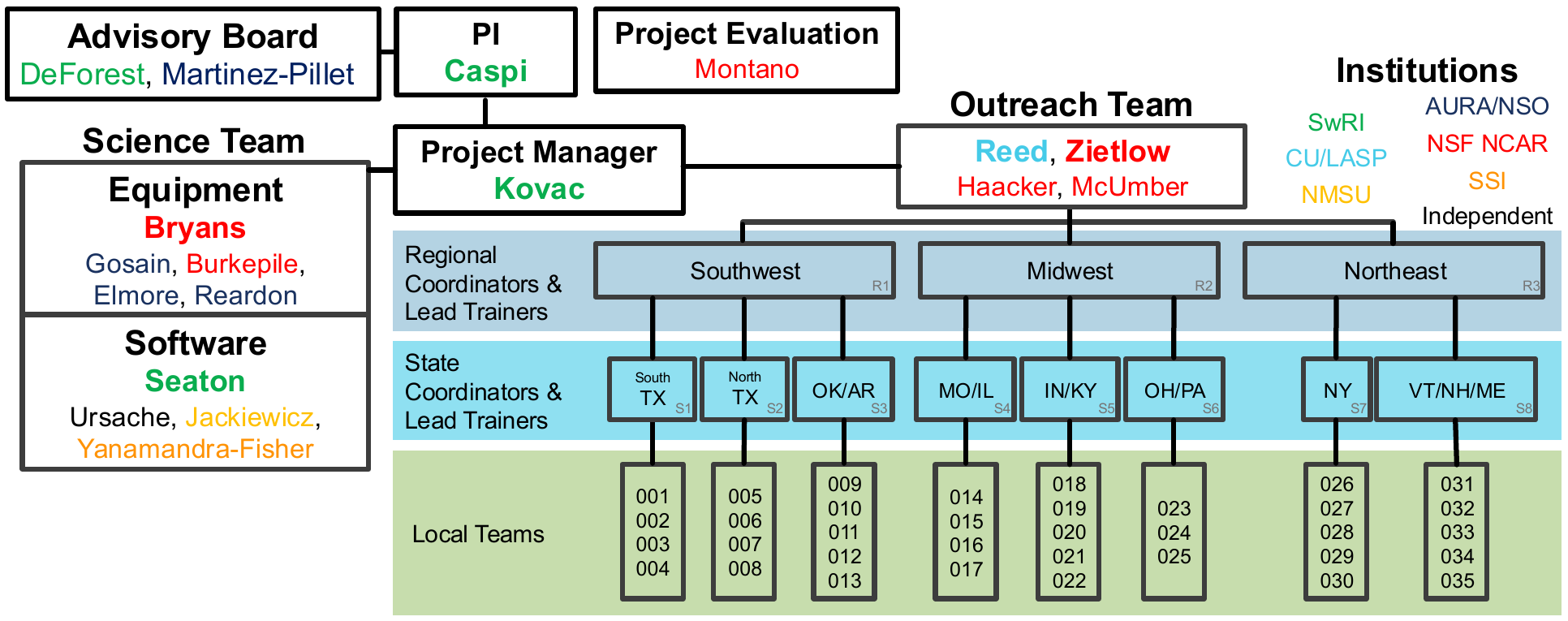}}
    \caption{CATE~2024 project personnel structure. This project was run by Southwest Research Institute (SwRI) along with partner institutions including the U.S. National Science Foundation National Center for Atmospheric Research (NSF NCAR), New Mexico State University (NMSU), the Association of Universities for Research in Astronomy National Solar Observatory (AURA/NSO), Space Science Institute (SSI), and the University of Colorado Laboratory for Atmospheric and Space Physics (CU/LASP).}
    \label{figure:personnel}
\end{figure}

Building on the experiences and lessons learned from the original 2017 Citizen CATE Experiment, the management of CATE~2024 was structured so that there were multiple levels of support such that no one team member would become overburdened. There are significant benefits of this multi-layer approach, including that there were multiple people whom folks could go to for help. At the top level, there was the core CATE~2024 team, who were reachable by all community participants, including Coordinators, team leads, and local team members. The core team member roles and affiliations are shown in Figure~\ref{figure:personnel}. The PI and Project Manager oversaw all aspects of the project, including the science and outreach teams, recruitment, procuring equipment, updating the observing procedure, and maintaining open lines of communication with all team members. The science team was responsible for testing and characterizing hardware, in addition to analyzing data from practice sessions and, ultimately, eclipse day. The outreach team focused on creating materials to distribute pre- and post-eclipse, as well as public engagement day-of. All CATE~2024 participants, from the core team to the individual team members, were required to sign a participation agreement that included a code of conduct. This was put in place for the safety and comfort of all team members.

\subsubsection{Regional- and State-Level Coordinators}
The roles of Regional- and State-Level Coordinators were primarily to help recruit and train local teams, so they needed to be well connected to their local communities. Some Coordinators split the role between two people to share responsibilities and further broaden engagement and local connections. Regional Coordinators helped recruit State Coordinators, and State Coordinators helped select local teams (with the support of Regional Coordinators). No prior astronomy or telescope experience was required for Coordinators, but it was vital that they were well ingrained in their communities. 

Since the Regional Coordinators needed to have deep and authentic connections along the path of totality, we relied on existing networks and the networks of individuals within those networks. Once selected, the Regional Coordinators then assisted the core CATE~2024 team with recruiting the State Coordinators. Regional Coordinators each oversaw 2--3~State Coordinators, and each State Coordinator oversaw 3--5~local teams. 

Recruitment of Coordinators occurred at professional conferences, public engagement events, and via social media.  All potential Coordinators were interviewed by the CATE~2024 Project Manager and outreach personnel. Topics discussed were their connections within their community (particularly with working with underserved populations), prior experience in scientific public engagement, and the time commitment needed for the role. Final selections were made by the project PI and Project Manager, with feedback from project team members with expertise in engagement and making sure the engagement goals of the project were met.

In total, three Regional Coordinators and 12~State Coordinators were recruited. While there were only eight state-level regions, some regions (e.g., Texas, where teams were spaced far apart) benefited from having two State Coordinators to reduce workload and improve network breadth. Background for each Coordinator is shown in Table~\ref{table:roles} and a group photo is shown in Figure~\ref{figure:group_photo}.

\subsubsection{Lead Trainers}
Regional and State Coordinators each selected a Lead Trainer to support their efforts in training participants. Oftentimes, this was a student from their local school, university, or community. While Coordinators were primarily responsible for recruitment of teams and helping with project logistics, the Lead Trainers supported the hands-on training of participants and became the equipment experts for their region. The CATE~2024 Project Manager met with each Lead Trainer to discuss roles and responsibilities. 

\begin{figure}
    \centering
    \centerline{\includegraphics[width=\textwidth]{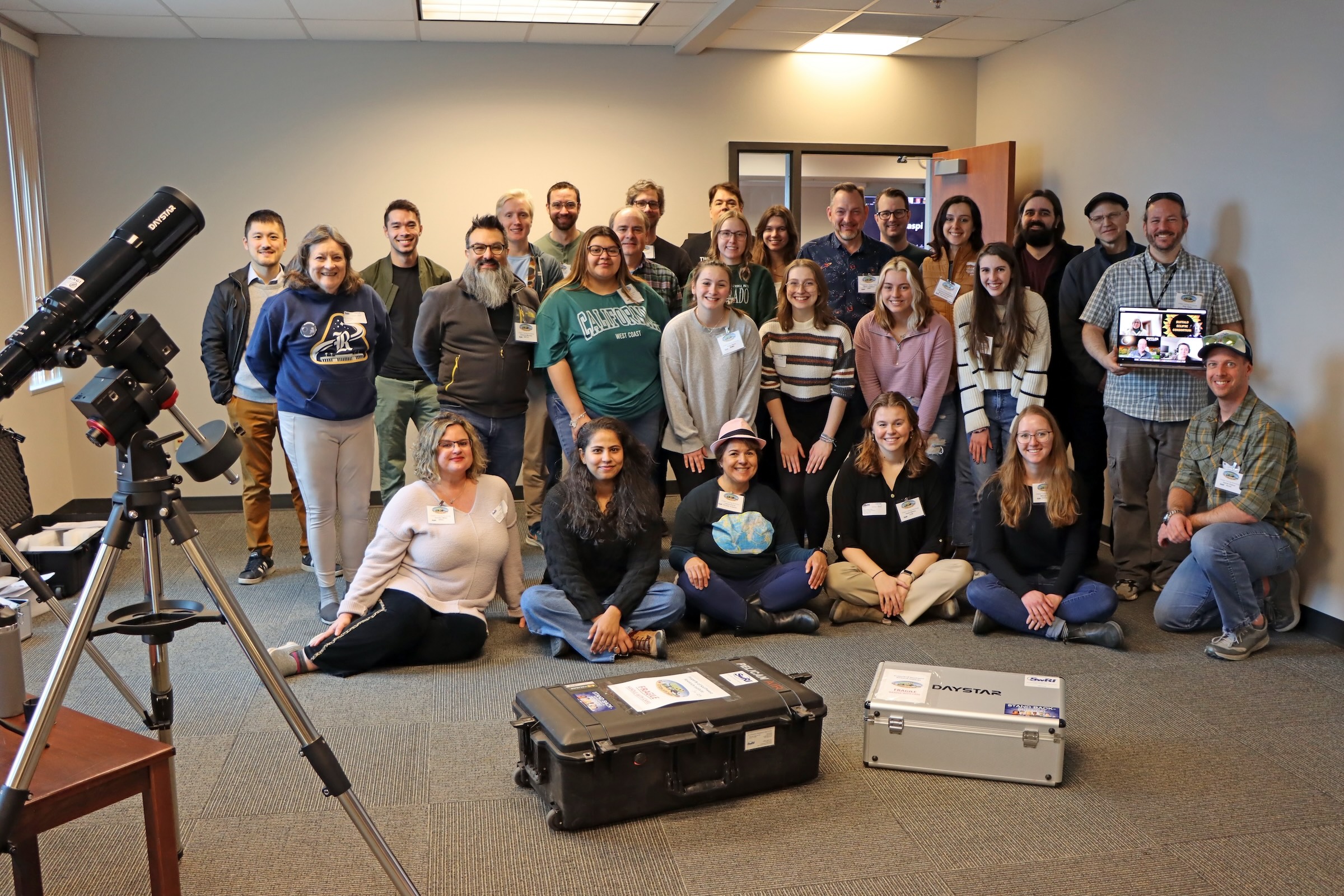}}
    \caption{Regional and State Coordinators, along with their Lead Trainers, gathered at the headquarters of SwRI in San Antonio,~TX to learn about their roles in CATE~2024, how to set up the telescopes, and how to engage their communities in the project, 19--21 January 2024.}
    \label{figure:group_photo}
\end{figure}

Lead Trainers often went out of their way to provide supplemental training to teams who needed extra support, including traveling to those teams on practice weekends. This gave undergraduate and graduate students the opportunity to experience a leadership role on a scientific project. Information for each trainer is shown in Table~\ref{table:roles} and a group photo is shown in Figure~\ref{figure:group_photo}.

\newpage
\begin{landscape}
\begin{table}[!ht]
    \caption{List of Coordinator and Trainer roles at the time of their participation in CATE~2024. RC~=~Regional Coordinator. SC~=~State Coordinator. LT~=~Lead Trainer.}
    \begin{tabular}{llll}
    \hline
        \textbf{Name} & \textbf{CATE 2024 Role} & \textbf{Affiliation} & \textbf{Career Level/Position} \\ \hline
        Patricia Reiff & RC -- Southwest & Rice Univ. & Prof. of Physics \& Astro. \\ 
        Charlie Gardner & Regional LT -- Southwest & Rice Univ. & Graduate Student \\ 
        John Carini & RC -- Midwest & Indiana Univ. Bloomington & Assoc. Prof. of Physics \\ 
        Rachael Weir & Regional LT -- Midwest & Indiana Univ. Bloomington & Undergraduate Student \\ 
        Shawn Laatsch & RC -- Northeast & Univ. of Maine & Planetarium Director \\ 
        Nikita Saini & Regional LT -- Northeast & Univ. of Maine & Graduate Student \\ 
        Sarah Davis & National LT & Univ. of N. Colorado & Physics Undergraduate Student \\ 
        Jennifer Miller-Ray & SC -- Southern TX & Sul Ross State Univ. & Assoc. Professor of Education\\ 
        Catarino Morales III & SC -- Southern TX & Southwest Texas Junior College & Biology Faculty \\ 
        Andrea Rivera & State LT -- Southern TX & Sul Ross State Univ. & Mathematics Undergraduate Student \\ 
        Yue Zhang & SC -- Northern TX & Texas A\&M Univ. & Asst. Prof. of Atmospheric Sciences \\ 
        Leticia Ferrer & SC -- Northern TX & Texas Solar Eclipses & Podcaster and Speaker \\ 
        Eden Thompson & State LT -- Northern TX & Texas A\&M Univ. & Meteorology Undergraduate Student \\ 
        Jeremy Lusk & SC -- AR/OK & Univ. of Central Arkansas & Asst. Prof. of Physics \\ 
        Erik Stinnett & State LT -- AR/OK & Univ. of Central Arkansas & Comp. Sci. Undergraduate Student \\ 
        Peggy Hill & SC -- MO/IL & Southeast Missouri State Univ. & Retd. Physics Prof. \\ 
        Jonathan Kessler & SC -- MO/IL & Southeast Missouri State Univ. & Assoc. Prof. of Physics \\ 
        Kayla Olson & State LT -- MO/IL & Southeast Missouri State Univ. & Environ. Sci. Undergraduate Student \\ 
        Brian Murphy & SC -- IN  & Butler Univ. & Prof. of Astronomy \\ 
        Aarran Shaw & SC -- IN  & Butler Univ. & Asst. Prof. of Astronomy \\ 
        Kira Baasch & State LT -- IN & Butler Univ. & Astrophysics Undergraduate Student \\ 
        Gwen Perry & SC -- OH/PA & \textit{unaffiliated} & ~ \\ 
        Ryan Ferko & State LT -- OH/PA & The Pennsylvania State Univ. & Alumnus \\ 
        Mark Percy & SC -- NY & Buffalo Eclipse Consortium & Founder \\ 
        Tim Collins & State LT -- NY & Buffalo Museum of Science & Research Associate  \\ 
        Jackie Bellefontaine & SC -- VT/NH/ME & Mount Washington Observatory & School Programs Coordinator \\ 
        Hazel Wilkins & State LT -- VT/NH/ME & Univ. of Vermont & Physics Undergraduate Student \\ 
        \hline
    \end{tabular}
\label{table:roles}
\end{table}
\end{landscape}
\newpage

\subsubsection{Local Teams}
The individual teams were the backbone of CATE~2024. Local teams comprised at least three individuals and were responsible for choosing an observing location, meaning they scouted viable observing locations throughout their local communities. All team members were tasked with attending training workshops, practice sessions, and conducting observations on eclipse day. 

The core CATE~2024 team developed and electronically distributed an interest form for community members who were interested in participating. Regional and State Coordinators worked to share the CATE~2024 interest form with various communities (teachers and schools, planetariums, museums, libraries, local clubs, etc.). The interest form was also posted to social media and shared widely. CATE~2024 received 379~statements of interest for 35~team slots. Even with the impressive number of total applications, some regions, like rural Maine, received only one, or even zero, applications. For these regions, the core CATE~2024 team worked with the Coordinators to personally invite teams or individuals to fill these spots, sometimes traveling from outside of the path of totality. 

Statements of interest were evaluated based primarily on participants' described reasons for wanting to be involved in the project and benefit to them. Since local teams needed to be on-site multiple times prior to the eclipse to scout potential locations and perform practice observations in the months leading up to the eclipse, we also prioritized teams who lived near their observing site and had familiarity with the geographic surroundings. Prioritizing choosing participants who lived along the path further increased the engagement within local communities as the teams could show their neighbors and families what they were up to with CATE~2024. Other evaluation criteria included availability to attend training workshops and practice sessions leading up to eclipse day, and connections within their community to house the telescopes after the eclipse.

\subsubsection{Site Location Selection}
The 35~participant-led CATE~2024 observing locations needed to be adequately spaced to capture 60~continuous minutes of totality. The top priority for site selection was to space teams such that there was dual coverage along the entire U.S. eclipse path. This helped mitigate the chances of having gaps in the data even if not all sites observed successfully.

\begin{figure}[!ht]
    \centering
    \centerline{\includegraphics[width=\textwidth]{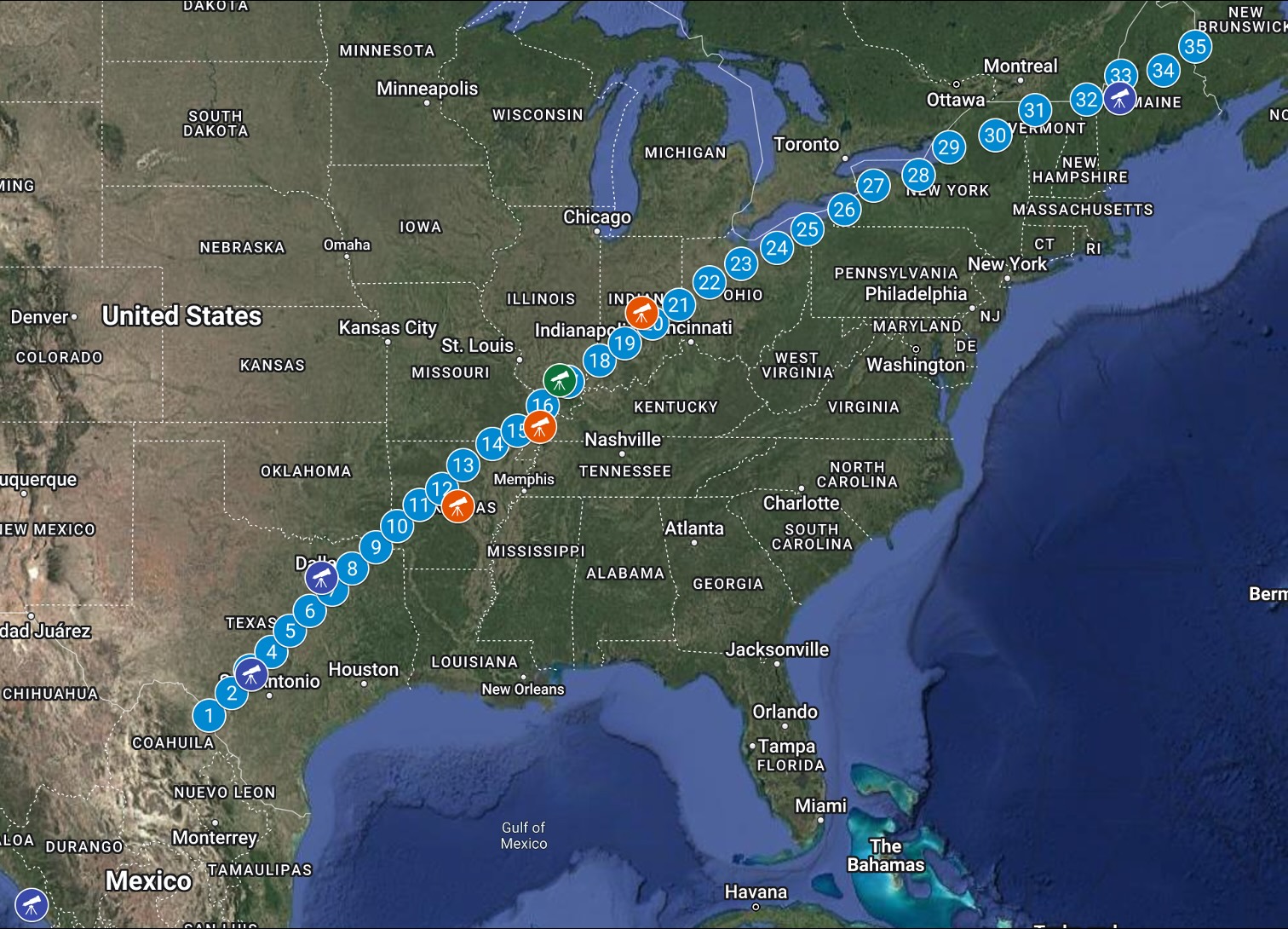}}
    \caption{CATE~2024 site locations. Light blue circles represent local community team locations. Orange circles represent locations of eclipse-day engagement sites. Green and dark blue circles represent additional engagement and professionally-led observing teams, respectively. Image made with Google Maps.}
    \label{figure:site_map}
\end{figure}

Once local teams were selected, we worked with them to find viable locations as close as possible to our ideal locations. Teams needed to stay within roughly five~miles of our ideally spaced location to ensure there was dual coverage along the path. In some regions, it was geographically impossible to get to the ideal location, or it posed a safety risk to teams. This was particularly true in Maine, where sites needed to be spaced further apart from ideal locations due to dense forests and mountains with limited access roads. We worked diligently with teams to ensure they chose safe observing locations. Secondary factors for teams were amenities such as power and restrooms. For remote sites without power, external battery packs were provided. Locations of each team on eclipse day are shown in Figure~\ref{figure:site_map} and detailed in Table~\ref{table:sites}.

In addition to  community participant-led sites, members of the CATE~2024 core team led observation and engagement sites at three dedicated outreach events (see Section~\ref{subsec:engagement}). Core team members also participated in the public NOAA/NSF/NASA event at the Cotton Bowl in Dallas,~TX, where they demonstrated equipment for event attendees, media, and agency representatives, and gathered eclipse observations for rapid release following the eclipse. The Cotton Bowl station was operated by a combination of core team members (Seaton and Caspi) and school-aged children from the greater Boulder area. A CATE~2024 team (Kovac and Davis) also participated in the NASA event in Kerrville,~TX, interacting with over 5,000~people on eclipse day. Despite being clouded out for totality, the Kerrville team provided the public with a unique view of the partial phases of the eclipse, and distributed various outreach materials. 

\begin{table}
\caption{List of all CATE~2024 observing locations. ``M'' sites refer to mobile sites, run by both amateurs and professionals. ``O'' sites refer to outreach-dedicated sites, run by outreach professionals.}
\label{table:sites}
\addtolength{\tabcolsep}{-.5em}
\begin{tabular}{lccccc}
\hline
CATE 2024 & Latitude & Longitude & Nearest & Totality Start & Totality End\\
Site Number & (degrees N) & (degrees W) & Town & Time (UTC) & Time (UTC) \\
\hline
R1-S1-001 & 28{\degree} 51$'$ 35$''$ & 100{\degree} 31$'$ 31$''$ & Eagle Pass, TX & 18:27:44 & 18:32:11 \\ 
R1-S1-002$^*$ & 29{\degree} 30$'$ 12$''$ & 99{\degree} 46$'$ 57$''$ & Uvalde, TX & 18:30:07 & 18:34:33 \\  
R1-S1-003$^*$ & 30{\degree} 8$'$ 23$''$ & 99{\degree} 10$'$ 40$''$ & Kerrville, TX & 18:32:13 & 18:36:39 \\
R1-S1-004 & 30{\degree} 40$'$ 27$''$ & 98{\degree} 28$'$ 46$''$ & Kingsland, TX & 18:34:17 & 18:38:42 \\ 
R1-S2-005 & 31{\degree} 16$'$ 46$''$ & 97{\degree} 53$'$ 10$''$ & Pidcoke, TX & 18:36:17 & 18:40:41\\ 
R1-S2-006 & 31{\degree} 51$'$ 22$''$ & 97{\degree} 13$'$ 6$''$ & Hillsboro, TX & 18:38:19 & 18:42:42 \\
R1-S2-007 & 32{\degree} 26$'$ 56$''$ & 96{\degree} 27$'$ 52$''$ & Kaufman, TX & 18:40:31 & 18:44:53 \\ 
R1-S2-008 & 33{\degree} 1$'$ 11$''$ & 95{\degree} 48$'$ 7$''$ & Miller Grove, TX & 18:42:29 & 18:46:50 \\
R1-S3-009 & 33{\degree} 37$'$ 48$''$ & 95{\degree} 0$'$ 44$''$ & Clarksville, TX & 18:44:41 & 18:49:01 \\
R1-S3-010 & 34{\degree} 10$'$ 35$''$ & 94{\degree} 19$'$ 12$''$ & DeQueen, AR & 18:46:37 & 18:50:56 \\
R1-S3-011 & 34{\degree} 47$'$ 2$''$ & 93{\degree} 33$'$ 1$''$ & Hot Springs, AR & 18:48:44 & 18:53:01 \\
R1-S3-012 & 35{\degree} 13$'$ 25$''$ & 92{\degree} 48$'$ 11$''$ & Blackwell, AR & 18:50:32 & 18:54:48 \\ 
R1-S3-013 & 35{\degree} 51$'$ 26$''$ & 92{\degree} 5$'$ 22$''$ & Mountain View, AR & 18:52:31 & 18:56:45  \\ 
R1-S3-014 & 36{\degree} 25$'$ 8$''$ & 91{\degree} 8$'$ 29$''$ & Dalton, AR & 18:54:44 & 18:58:56\\ 
R2-S4-015 & 36{\degree} 45$'$ 55$''$ & 90{\degree} 19$'$ 21$''$ & Poplar Bluff, MO & 18:56:29 & 19:00:35 \\ 
R2-S4-016 & 37{\degree} 26$'$ 32$''$ & 89{\degree} 29$'$ 39$''$ & Cape Girardeau, MO & 18:58:32 & 19:02:41 \\ 
R2-S4-017 & 38{\degree} 4$'$ 57$''$ & 88{\degree} 37$'$ 31$''$ & McLeansboro, IL & 19:00:35 & 19:04:42 \\ 
R2-S5-018 & 38{\degree} 36$'$ 34$''$ & 87{\degree} 35$'$ 29$''$ & St. Francisville, IL & 19:02:43 & 19:06:48 \\ 
R2-S5-019 & 39{\degree} 2$'$ 60$''$ & 86{\degree} 45$'$ 14$''$ & Bloomington, IN & 19:04:25 & 19:08:28 \\ 
R2-S5-020 & 39{\degree} 35$'$ 3$''$ & 85{\degree} 48$'$ 39$''$ & Fairland, IN & 19:06:20 & 19:10:20 \\ 
R2-S5-021 & 40{\degree} 3$'$ 5$''$ & 84{\degree} 56$'$ 40$''$ & Crete, IN & 19:08:01 & 19:12:00 \\ 
R2-S5-022 & 40{\degree} 35$'$ 52$''$ & 83{\degree} 56$'$ 38$''$ & Waynesfield, OH & 19:09:54 & 19:13:51 \\ 
R2-S6-023 & 41{\degree} 4$'$ 0$''$ & 82{\degree} 53$'$ 45$''$ & Attica, OH & 19:11:45 & 19:15:40 \\ 
R2-S6-024 & 41{\degree} 29$'$ 8$''$ & 81{\degree} 42$'$ 47$''$ & Cleveland, OH & 19:13:44 & 19:17:33 \\ 
R2-S6-025 & 41{\degree} 55$'$ 7$''$ & 80{\degree} 41$'$ 25$''$ & Ashtabula, OH & 19:15:26 & 19:19:11 \\ 
R3-S7-026$^*$ & 42{\degree} 25$'$ 33$''$ & 79{\degree} 25$'$ 56$''$ & Fredonia, NY & 19:17:25 & 19:21:07 \\ 
R3-S7-027$^*$ & 43{\degree} 0$'$ 52$''$ & 78{\degree} 29$'$ 35$''$ & Akron, NY & 19:18:55 & 19:22:40 \\ 
R3-S7-028$^*$ & 43{\degree} 16$'$ 15$''$ & 76{\degree} 58$'$ 53$''$ & Sodus Point, NY & 19:21:05 & 19:24:34 \\ 
R3-S7-029 & 43{\degree} 55$'$ 20$''$ & 75{\degree} 58$'$ 35$''$ & Watertown, NY & 19:22:29 & 19:26:08 \\ 
R3-S8-030 & 44{\degree} 13$'$ 31$''$ & 74{\degree} 26$'$ 28$''$ & Tupper Lake, NY & 19:24:31 & 19:28:00 \\ 
R3-S8-031 & 44{\degree} 48$'$ 32$''$ & 73{\degree} 8$'$ 32$''$ & St. Albans, VT & 19:26:08 & 19:29:41 \\ 
R3-S8-032 & 45{\degree} 3$'$ 11$''$ & 71{\degree} 23$'$ 28$''$ & Pittsburg, NH & 19:28:19 & 19:31:34 \\ 
R3-S8-033 & 45{\degree} 37$'$ 15$''$ & 70{\degree} 14$'$ 53$''$ & Jackman, ME & 19:29:31 & 19:32:58 \\ 
R3-S8-034 & 45{\degree} 43$'$ 37$''$ & 68{\degree} 51$'$ 45$''$ & Millinocket, ME & 19:31:08 & 19:34:17 \\ 
R3-S8-035 & 46{\degree} 17$'$ 28$''$ & 67{\degree} 48$'$ 50$''$ & Houlton, ME & 19:32:06 & 19:35:28 \\ 
\hline
M1-DE-036 & 23{\degree} 11$'$ 2$''$ & 106{\degree} 25$'$ 33$''$ & Mazatl\'{a}n, MX & 18:07:21 & 18:11:41 \\ 
M2-AF-037 & 38{\degree} 5$'$ 37$''$ & 88{\degree} 54$'$ 56$''$ & Benton, IL & 19:00:12 & 19:04:14 \\ 
M3-AU-038 & 45{\degree} 4$'$ 44$''$ & 70{\degree} 17$'$ 40$''$ & Carrabassett Valley, ME & 19:29:55 & 19:32:23\\ 
\hline
O1-TX-039$^*$ & 30{\degree} 2$'$ 45.$''$ & 99{\degree} 8$'$ 40$''$ & Kerrville, TX & 18:32:07 & 18:36:32 \\
O2-TX-040 & 32{\degree} 46$'$ 48$''$ & 96{\degree} 45$'$ 37$''$ & Dallas, TX & 18:40:46 & 18:44:40 \\
O3-AR-041 & 34{\degree} 44$'$ 6$''$ & 92{\degree} 16$'$ 21$''$ & Little Rock, AR & 18:51:43 & 18:54:01 \\
O4-MO-042 & 36{\degree} 53$'$ 26$''$ & 89{\degree} 34$'$ 4$''$ & Sikeston, MO & 18:58:07 & 19:01:37 \\
O5-IN-043 & 39{\degree} 50$'$ 28$''$ & 86{\degree} 10' 17$''$ & Indianapolis, IN & 19:06:11 & 19:09:53 \\
\hline
\end{tabular}
\begin{tablenotes}
   \item[*]$^*${These sites did not observe totality due to poor weather conditions.}
\end{tablenotes}
\end{table}

\subsubsection{Public Communications}
A broad communications plan was established to ensure that key objectives of the project were communicated to a wide audience. The plan identified different types of audiences we wanted to reach and how to curate messaging to engage those audiences. Resources were shared with all project members, including training material on how to engage with the media and general public. These tools were highly beneficial in ensuring CATE~2024 project members at all levels were communicating the correct objectives for every phase of the project.

The key themes we identified to communicate with all audiences were:
\begin{enumerate}
    \item Polarization is important in solar science and studying polarization is what makes CATE~2024 unique compared to other projects.
    \item CATE~2024 will have a wide reach, be approachable and inclusive, and work to bring solar science to communities.
    \item A large team with different experience is central to the success of CATE~2024, meaning anyone can be a part of the project.
    \item The total eclipse is a community bonding experience and the CATE~2024 project is extending this feeling across the entire path of totality.
\end{enumerate}

These themes were emphasized in all products that the project produced, including the website, recruitment flyers and emails, social media posts, educational materials, and press materials. 

Social media was an important element in engaging the public about the project, especially in the days around eclipse day. In particular, CATE~2024 utilized three social media platforms: Facebook, Instagram, and X, using the username @CitizenCATE2024. According to platform analytics, during the week of the eclipse, several thousand social media users saw our posts on Facebook, Instagram, and X. This includes direct interaction with the posts via re-sharing of posts, writing comments on the posts, and likes. Some team members also shared their own individual posts on these platforms, or other platforms such as BlueSky, for even broader engagement.

\subsection{Instrumentation}\label{sec:pi_instrument}
The instrumentation chosen for CATE~2024 expands the setup developed by the original 2017 Citizen CATE Experiment, using it as a baseline but changing several pieces of equipment, most notably the camera. The new instrumentation choices, shown in Figure~\ref{figure:schematic} and detailed in Table~\ref{table:equipment}, were driven by the observational requirements needed to meet the 2024 science goals, described in Table~\ref{table:performance_requirements}, and result from an engineering run during the 2023~TSE (see Section~\ref{sec:2023}).

\subsubsection{Camera and Telescope}
CATE~2024 teams used the FLIR BFS-PGE-123S6P-C Blackfly camera with next-generation Sony IMX253/MZR 12MP CMOS~sensor. This sensor has a $4096 \times 3000$~pixel array with 3.45~$\mu$m pixel pitch and integrated polarizer mask providing simultaneous polarization measurements in four angles ($0^{\circ}$, $45^{\circ}$, $90^{\circ}$, $135^{\circ}$), enabling recovery of Stokes I, Q, and U across the entire field of view. The camera provides a GigE interface compatible with all modern laptops and includes features useful for CATE~2024’s high-speed HDR sequencing. The camera captured images through a Long Perng S500G-A refractor telescope, procured via Daystar, with an 80~mm aperture and 500~mm focal length (\textit{f}/6.25) and an ED APO S-FPL53 doublet, fully multi-coated lens. Together, the camera and telescope combination provide a $1.43''$/pixel platescale and $1.627^{\circ} \times 1.192^{\circ}$ field of view.

\subsubsection{Telescope Mount}
An iOptron GEM28 German Equatorial GoTo Mount was used to drive the telescope and included a tripod and 5~kg counterweight. CATE~2024 opted to upgrade to a go-to mount to make it more likely we would reach our engagement goals, as these mounts are much more user-friendly for nighttime observing. While there was a steeper learning curve, go-to mounts are highly preferred for most observations, making it more likely the telescope would be used by communities post-eclipse. While the mount included a camera-based polar scope (iPolar) for nighttime alignment, it was not used on eclipse day.

For teams at lower latitudes, as the telescope tracked throughout the day, the position of the 5~kg counterweight caused it to impact the mount. Observing locations below $32^{\circ}$~latitude (Sites~001--006) were provided a supplemental smaller 2~kg counterweight. Being lighter, the 2~kg counterweight sat farther on the counterweight shaft, eliminating the risk of impact.

\subsubsection{Laptop, Software, and GPS}
The laptop needed to be able to run the custom CATE~2024 observing software and have enough storage for the necessary calibration and totality data. Equally important, the machine needed to process data fast enough that we had real-time image display while observing. The Acer Aspire Vero AV14-51-73LM, including an Intel Core~i7 processor with 16~GB RAM and 1--2~TB SSD storage, was field-tested to have all of the capabilities required while staying relatively low cost.

The custom MATLAB software used in the 2017 Citizen CATE Experiment was redesigned to fit the needs of CATE~2024. This included changes to the user interface to make the experience more user-friendly. The laptop was connected to a VK-162 USB-based GPS dongle. The GPS gave precise location information including latitude, longitude, and altitude, and provided an absolute time reference for image acquisition.

\subsubsection{Additional Optics}
On eclipse day, teams used a Daystar solar finder (a simple but precisely-machined pinhole projection onto a crosshair target) and solar filter. The telescope was equipped with a Daystar DSIUV2 UV/IR cut filter to reduce stray light. For teams to take robust flat-field images, a 10~cm $\times$ 10~cm broadband hybrid diffuser glass from Edmund Optics (model~36-618) was provided with a custom 3D-printed holder. Broadband hybrid diffusers are semi-opaque and specifically designed to highly scatter light from the UV to the near-IR. 

\begin{figure}[!ht]
    \centering
    \centerline{\includegraphics[width=\textwidth]{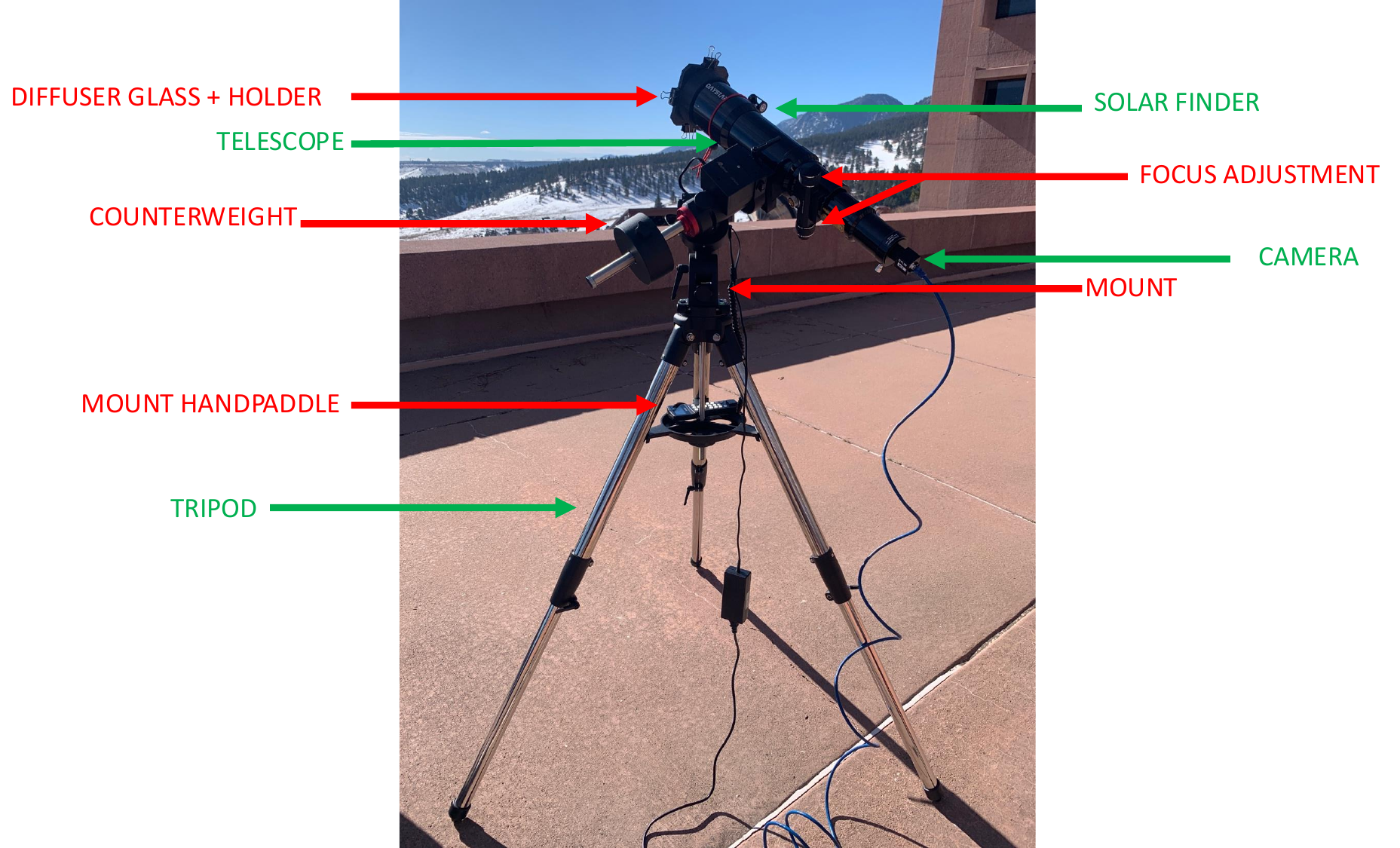}}
    \caption{Schematic of CATE~2024 setup with the equipment that was used on eclipse day.}
    \label{figure:schematic}
\end{figure}

CATE~2024 teams were provided additional equipment which was not used during the eclipse and was intended for post-eclipse activities. This included an optical field flattener to aid in astrophotography, as well as multiple eyepieces for casual telescope use. A full list of equipment can be found in Table~\ref{table:equipment}. CATE~2024 also distributed over 20,000~solar viewing glasses generously provided by NASA, the American Astronomical Society, and NSF NCAR. We also distributed several hundred pinhole projector cards, in both English and Spanish, created by the NASA \textit{PUNCH} outreach team,\footnote{\url{https://punch.space.swri.edu/punch_outreach_pinholeprojector.php}} and NSF NCAR distributed additional informational cards.

\begin{table}
\caption{List of CATE~2024 station equipment. The top section lists all equipment that was used on eclipse day. The bottom section lists supplemental equipment not needed on eclipse day. The items in italics were distributed on an as-needed basis; not all teams received these items.}
\label{table:equipment}
\begin{tabular}{ll}
\hline
Item & Manufacturer \& Model Number \\
\hline
Camera & FLIR BFS-PGE-123S6P-C; 12~MP, polarization \\
Telescope & Long Perng S500G-A (via Daystar); 80~mm (\textit{f}/6.25) \\
Mount & iOptron GEM28 German Equatorial GoTo Mount; \\
~ & included iPolar, tripod, and 5~kg counterweight \\
Laptop & Aspire Vero AV14-51-73LM; Intel Core~i7 processor, \\
~ & 16~GB RAM, 1--2~TB SSD Storage \\
UV/IR Cut Filter & Daystar DSIUV2; 2'' diameter\\
Solar Filter & Daystar WLF90; 4.25'' full aperture \\
Solar Finder & Daystar USF \\
Diffuser & Edmund Optics 36-618; 10~cm $\times$ 10~cm \\
Diffuser Holder & SwRI (3D-print)$^1$ \\
Thumb Drive & Samsung MUF-256DA/AM; 256GB \\
Shipping Case & Pelican Air 1615 \\
Locks for Pelican & Masterlock 4696T (TSA-compliant) \\
Power-over-Ethernet Adapter & TRENDnet TPE-113GI \\
USB-to-Ethernet Adapter & Anker AK-A83130A1 \\
Ethernet Cable & GearIT CAT6E (qty: 2) \\
GPS & VK-162 \\
Optical Adapters & Various brands; 2'' 80~mm extension tube; \\
~ & 2'' to C-mount; T42 variable spacers; T42 to C-mount;  \\
~ & C-mount threaded cap \\
\hline
Field Flattener & Long Perng F1.0X-AA (via Daystar)\\
Diagonal Mirror & SVBONY SV188P\\
Eyepieces & SVBONY; 40~mm, 30~mm, 6.3~mm, 2$\times$ Barlow \\
9'' Torpedo Level & Newark \\
Bubble Levels & Various brands; 32~mm, 10~mm (qty: 2) \\
String & Various brands \\
Compass & Brunton Truarc 15 Luminous \\
\textit{External Power Source$^2$} & \textit{Jackery Explorer 300 Plus} \\
\textit{Counterweight} & \textit{iOptron; 2~kg} \\
\hline
\end{tabular}
\begin{tablenotes}
   \item[*]$^1${\url{https://www.tinkercad.com/things/bkvEqK36zIt-telescope-bracket-v2-all}}
   \item[*]$^2${Team~003 used a FIRMAN~P01204 generator as an external power source.}
  \end{tablenotes}
\end{table}

\subsubsection{Equipment Packaging}
As many teams would be traveling to remote locations on eclipse day, it was vital that the CATE~2024 setup would be easily portable. Packaging would also need to be robust, to protect the equipment during transit including shipping, and for any post-eclipse air travel (see Figure~\ref{fig:equipment}). The Pelican Air~1615 was chosen as it is regulation size for air travel, and all pieces of equipment needed for eclipse day fit into this one case. CATE~2024 teams also received the box the telescope originally came in, modified to hold extra equipment like eyepieces, field flattener, etc.

\begin{figure}
\centering
\addtocounter{figure}{-1}
\begin{subfigure}
  \centering
\includegraphics[width=1.0\linewidth,trim={0 0 0 80}, clip]{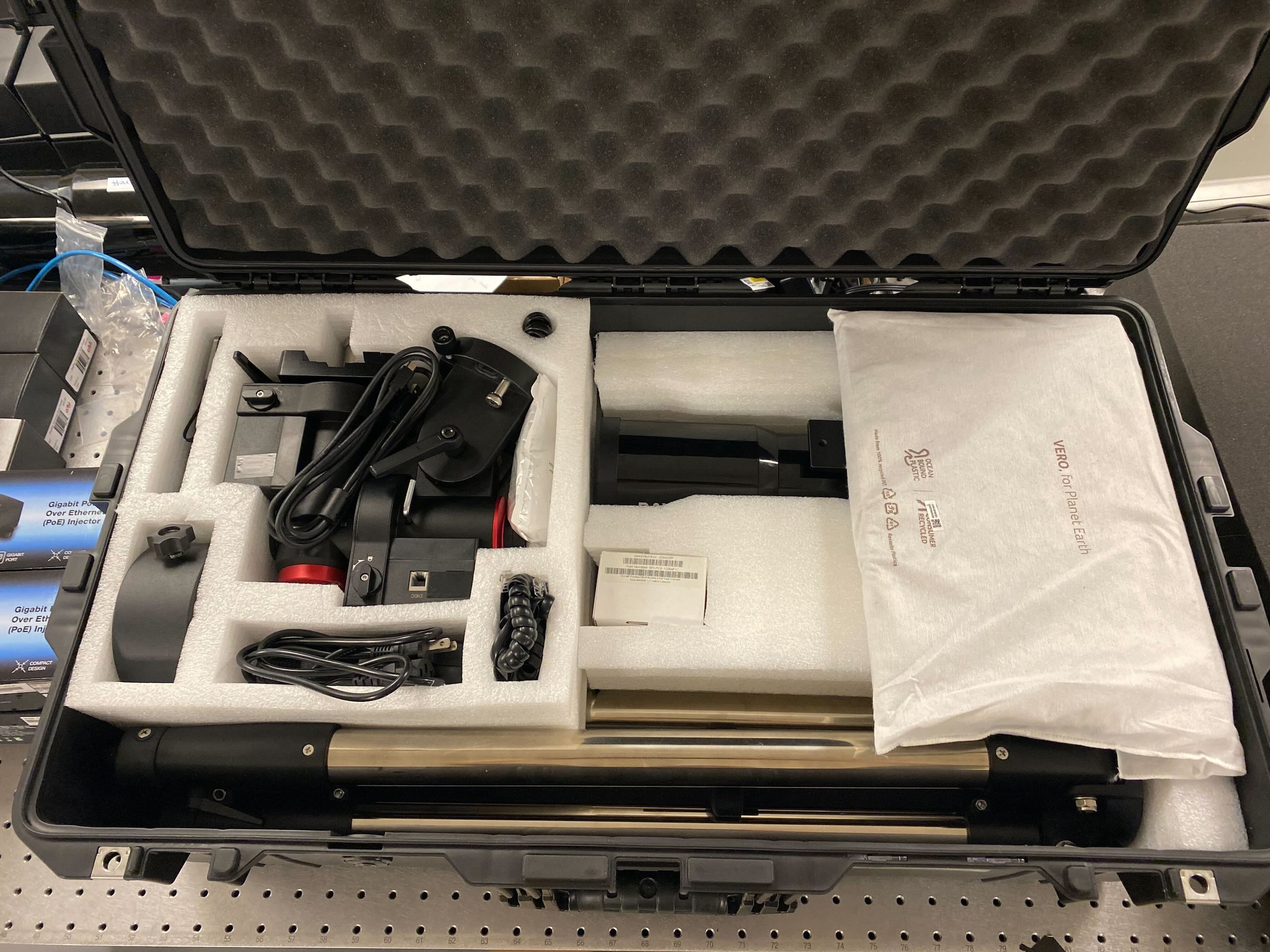}
\caption{Packed Pelican Air 1615}
  \label{fig:sub1}
\end{subfigure}
\vspace{1ex}
\addtocounter{figure}{-1}
\begin{subfigure}
  \centering
 \includegraphics[height=1.0\linewidth,trim={30 20 90 25},clip, angle=90]{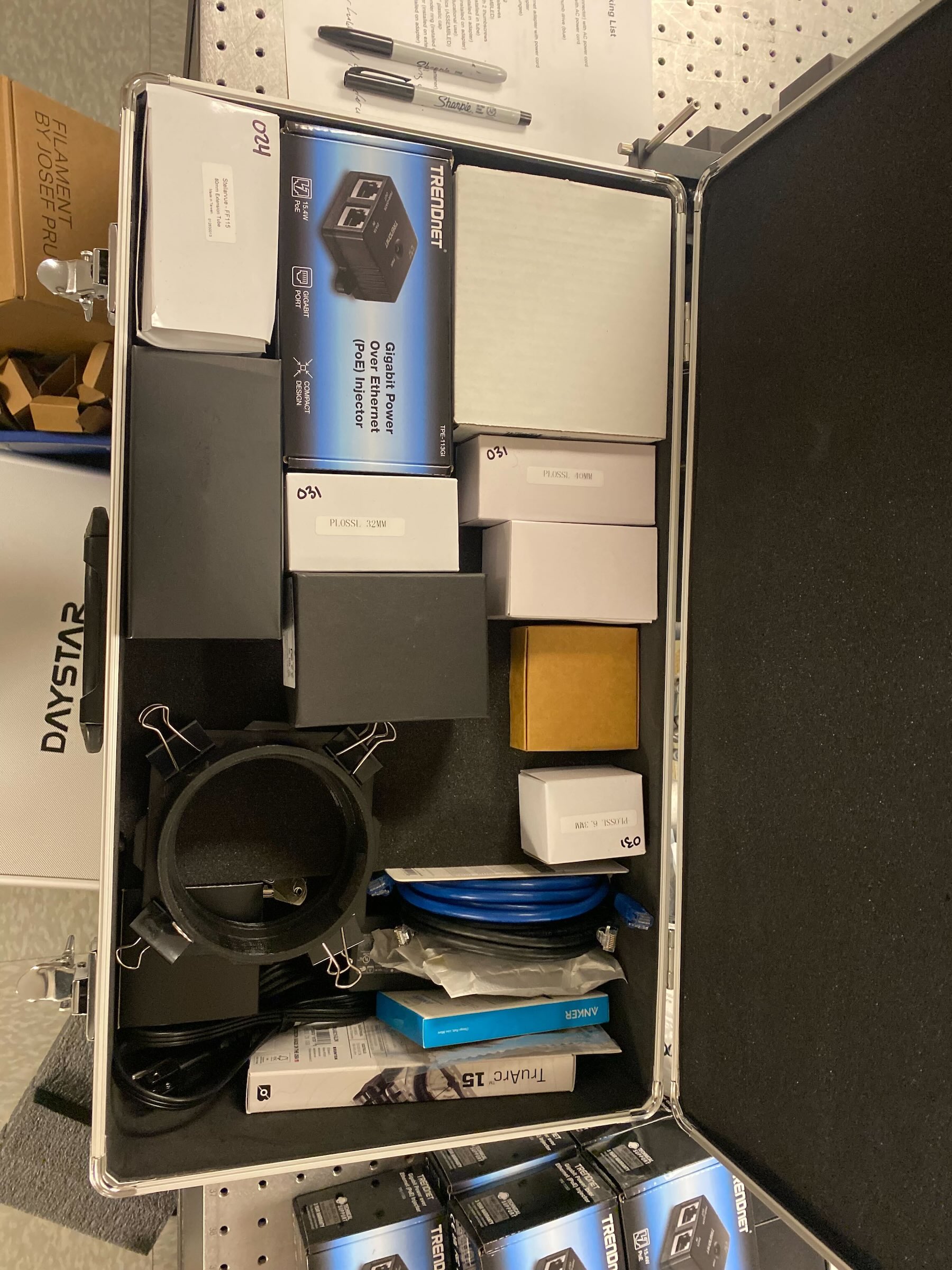}
\caption{Packed telescope box}
  \label{fig:sub2}
\end{subfigure}
\caption{Example of how equipment was packaged prior to shipment. (Top) Pelican Air~1615 housing the telescope, mount, tripod, camera, and laptop. (Bottom) Daystar case housing all supplemental equipment that did not fit into the Pelican case.}
\label{fig:equipment}
\end{figure}

\section{Observing Procedure}\label{sec:procedure}
The CATE~2024 observing procedure is a modified version of the procedures developed during the 2017 Citizen CATE Experiment \citep{Penn2020} and follow-on observations during the 2019~TSE \citep[]{cate_obs_2019, christensen2019}, updated to reflect the new instrumentation and lessons learned to improve understanding and ease of implementation. The procedure was designed to be executable by someone who had no previous experience with a telescope or data collection methods. A first draft of the 2024~procedure was created by the core team after procuring equipment. The procedure then underwent many iterations based on feedback from our community participants, first via the Regional Coordinators and Lead Trainers after the 2023~training workshop, again after the 2023~Australian eclipse expedition, and then several more times after getting feedback from local teams. The full procedure took roughly two hours to complete using the custom software developed in MATLAB. The software contained tabs that chronologically followed the written procedure: Configuration, GPS, Alignment (Figure~\ref{figure:software_alignment}), Focus (Figure~\ref{figure:software_focus}), Drift, and Totality. The Test tab was used for troubleshooting and debugging the software and was only used prior to eclipse day.

\subsection{Station Setup}
Each team started the observing procedure by setting up the tripod and roughly aligning to magnetic North. Teams then attached the mount and used a modified version of the Rackley method \citep{rackley2017} to more accurately polar align. Obtaining an accurate polar alignment without the use of multiple stars is notoriously difficult, so teams were provided with a compass to aid in the polar alignment process. Teams then inserted the telescope and camera, with all needed cords for power and data transfer attached, and the telescope was balanced.  

With the camera plugged into the computer, teams started the Solar Eclipse App and configured their site name using the Configuration tab. Teams were equipped with a solar finder for easy alignment to the Sun. The GPS was plugged in via USB and took a few minutes to calibrate. Teams took GPS snapshots via the GPS tab at least once per hour so that the drift of the laptop clock relative to the GPS absolute time reference could be calculated and calibrated out in post-processing. 

\subsection{Mount Settings}
CATE~2024 used a go-to mount, which requires the user to input geographic information so that the mount knows where it is, and therefore knows where to point. Teams then input date, latitude, longitude, current local time, and UTC~offset, and select a setting for daylight saving time (DST). Through experimentation, it was found that being off by even a few tens of seconds could result in tracking poorly enough that it would impact image quality. Teams were instructed to track at the solar rate, and were provided with a spreadsheet created by the Project Manager that outlined the inputs they needed to enter on eclipse day, specific to their location. 

Altitude limits prevent the telescope from tracking past a certain angle in the sky. If this limit is set higher than the altitude of the object being observed, the telescope will not be able to slew to the object. This adjustable setting needed to be set to zero degrees to ensure the telescope would track through its full range. If the telescope tracked indefinitely, the telescope would eventually hit the tripod, potentially damaging equipment. A meridian flip occurs when a target object is being observed as it crosses the meridian, causing the telescope to rotate to the other side of the tripod. Depending on the site location, teams also needed to adjust the meridian treatment angle such that the mount did not perform a meridian flip during totality. 

\subsection{Alignment}\label{sec:alignment}
\begin{figure}
    \centering
    \centerline{\includegraphics[width=\textwidth]{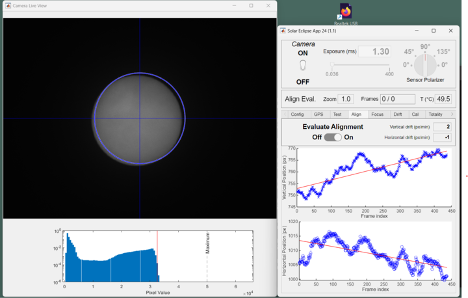}}
    \caption{Snapshot of the software CATE~2024 teams used during the 8~April 2024~TSE. This shows the Alignment tab.}
    \label{figure:software_alignment}
\end{figure}

The Solar Eclipse App included real-time feedback on alignment. Teams adjusted the exposure time of the camera such that no pixels were saturated in  the image (see Figure~\ref{figure:software_alignment}) and to center the Sun within the frame. The app included visual guides to help center the Sun. With the camera properly aligned, the image of the Sun moved only vertically when declination (Dec) was adjusted, and moved only horizontally when right ascension (RA) was adjusted.

Two sets of drift calibration images were collected to characterize the amount of tracking drift present and the accuracy of the camera alignment. Polar alignment was assessed by taking a series of 100~images (one per second). With solar tracking on and good polar alignment, the Sun should not drift from its initial position. The difference in the location of the Sun in the first image compared to the 100th~image quantified how much drift occurred. Camera alignment was assessed based on a series of 100~images (one per second) while the solar tracking was turned off. While the mount is not tracking, the Sun should drift significantly in RA, with negligible drift in Dec. The angle of the location of the Sun in the first image compared to that in the 100th~image, with perfect alignment, would be zero degrees relative to camera horizontal. 

\subsection{Focus}\label{sec:focus}
\begin{figure}
    \centering
    \centerline{\includegraphics[width=\textwidth]{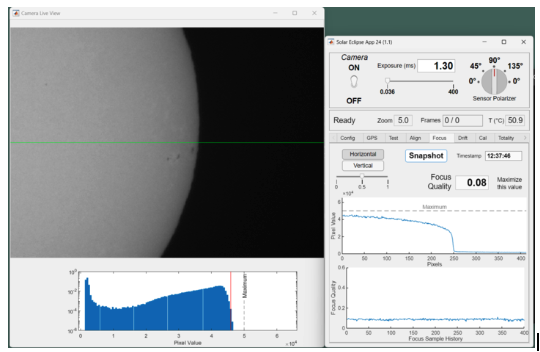}}
    \caption{Snapshot of the software CATE~2024 teams used during the 8~April 2024~TSE. This shows the Focus tab.}
    \label{figure:software_focus}
\end{figure}

Teams used the Focus tab to quantify the sharpness of the image. This was best done using the lunar limb after first contact. Prior to first contact, and during practice sessions, teams used sunspots to optimize focus. Figure~\ref{figure:software_focus} shows an example of the focus quality value across the solar limb. Focus quality was determined by the sharpness of the focus image snapshots. The sharp drop in intensity is consistent with well focused images. Once teams were satisfied with their focus, the telescope focus knob was manually locked. Since teams set up several hours before totality, the focus was adjusted over time to account for changes in temperature and seeing. 

\subsection{Calibration Data}
Calibration and other ancillary data (including GPS time and location) are used to normalize images and compute world coordinate system (WCS) metadata to be able to co-align images between sites. Darks were taken with the telescope dust cover attached using the same sequence of exposure times as the totality sequence, for a total of 20~sequences resulting in 160~dark frames. Flats were taken with the telescope pointed at the Sun with the broadband hybrid diffuser glass. Ideally, teams would collect this data prior to totality, but if they were unable, these data could be collected after third contact.

\subsection{Totality}
Teams finalized their focus roughly 20~minutes before totality and left the setup undisturbed for the remaining time leading up to totality. Teams started logging data about 10~seconds before totality actually started to ensure they caught the very first few seconds of totality, including Baily's beads at second contact. The app displayed the images of totality in real time, such that teams could keep an eye on the data stream to ensure it was being collected properly, and that the Sun was staying centered in frame. The series of eight exposures took roughly 1.6~seconds. With an average totality duration of four minutes, each team collected 1500--2000~individual totality frames (resulting in over 200~HDR images). Teams could then utilize the Quick Look tab, allowing them to create preliminary, non-calibrated HDRs immediately after totality. 

\subsection{Data Backup and Distribution}
Immediately following totality, teams backed up their data onto an external flash drive. Teams were instructed to keep this flash drive in a separate location than the laptop, such that if the Pelican case with the laptop were lost or damaged, there would be a backup copy of the data elsewhere. Teams were provided USPS shipping labels to mail the flash drives to the core team in Boulder,~CO. Additionally, a data repository was set up on Google Drive for teams to store a copy of their observations on the cloud. For some teams at sites with readily available internet, this was immediately following totality. For most teams, data was uploaded within a day or two after returning home from observing.

\section{Training Participants}\label{sec:training}
\subsection{2023 Eclipse Observations}\label{sec:2023}
The 20~April 2023~TSE across Western Australia and Southeast Asia provided an optimal opportunity prior to the 2024~TSE for beta-testing the CATE~2024 equipment, testing and enhancing the observing procedures, and providing both real field experience for team members and real polarized eclipse images to help develop the data processing and analysis pipeline for 2024. The 2023~TSE expedition comprised ten CATE~2024 team members, including three members from the CATE~2024 core team (Caspi, Klein, Zietlow), the National Lead Trainer (Davis), and the six Regional Coordinators and Lead Trainers (see Table~\ref{table:roles}). We leveraged this opportunity to test three potential configurations of the observing equipment to determine the optimal setup for mass distribution in 2024 based on performance and ease of use. The expedition also helped to refine both the observing software and the procedures, to improve the experience and chances for success in 2024.

In preparation for the 2023~TSE, a training workshop was held on 1--2~April 2023 in Boulder, CO. Efficiently training the Regional Coordinators and Lead Trainers was critical, as they would go on to help train the State Coordinators and local team participants. Training consisted of hands-on practice and guided presentations where attendees learned the history of the original 2017 Citizen CATE Experiment, an overview of the CATE~2024 next-generation project, the goals of the project and why they were important, plus general telescope operations and eclipse safety. A vast majority of the time, attendees were outside getting hands-on experience with the CATE~2024 equipment and becoming familiar with the observing procedures and software. 

The 10-person team embarked on the 2023~expedition on 12--13~April~2023, converging from across the country to meet in Los~Angeles en route to their final destination of Exmouth, Western Australia (via Melbourne, Perth, and Learmonth). Despite numerous flight delays, the team arrived safely in Exmouth on time, on 15~April~2023. While not directly on the eclipse centerline, Exmouth offered both convenient lodging and a safe and secure location with amenities including power to facilitate pre-eclipse practice and event-day observing, at the cost of losing only a few seconds of totality compared to a more remote and less accessible centerline location.

\begin{figure}[H]
    \centering
    \centerline{\includegraphics[width=\textwidth]{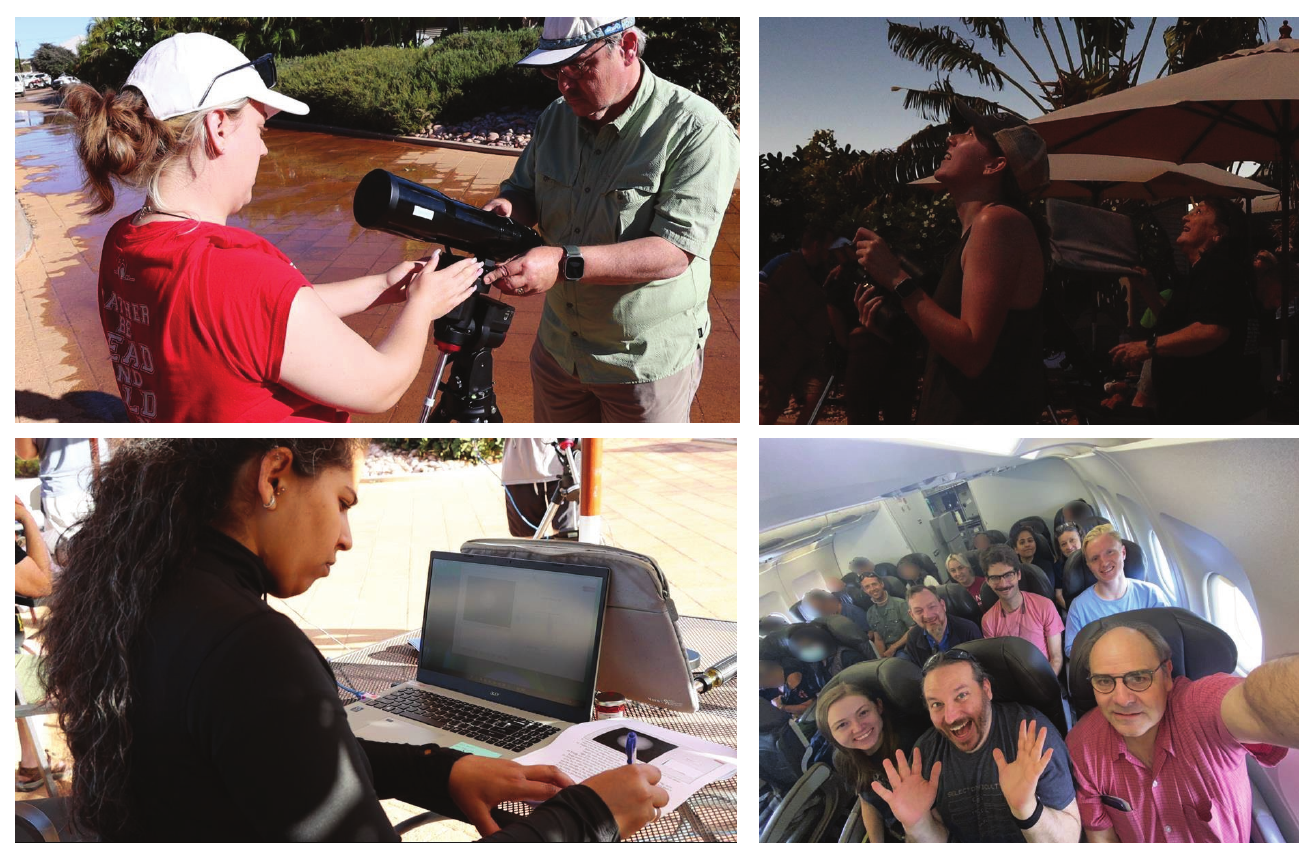}}
    \caption{Photographs from the 2023~TSE expedition in Exmouth, WA, AUS. (top left) R.~Weir and J.~Carini practicing with CATE~2024 hardware; (top right) S.~Davis and P.~Reiff experiencing totality; (bottom left) N.~Saini taking notes on the procedure while doing practice observations; (bottom right) 10-person team en route to Exmouth, WA, AUS; left to right, top to bottom: D.~Zietlow, R.~Weir, N.~Saini, P.~Reiff, S.~Laatsch, V.~Klein, C.~Gardner, S.~Davis, A.~Caspi, J.~Carini. Personnel roles are described in Table~\ref{table:roles}.}
    \label{figure:aus_people}
\end{figure}

Each of the three pairs of Regional Coordinators and their Lead Trainers were assigned an observing kit and spent the next few days practicing the setup and observing sequence, under the direction of the core CATE~2024 team and the National Lead Trainer, culminating in a dry-run the day before the eclipse. Each day's practice was followed by a discussion of issues encountered, solutions found, and lessons learned to inform the plan for the subsequent day and to continually improve the observing procedures. On the day of the eclipse, setup and observing proceeded nearly flawlessly, with all three teams making excellent, high-quality observations of 58~seconds of totality. Sample data were immediately uploaded over satellite internet to colleagues in the U.S. for processing and public release the next morning, U.S. time \citep{patel2023}. After necessary celebrations and data backup, the team repatriated to the U.S., arriving in Los~Angeles on 22~April~2023 (which, thanks to the international date line, was actually before they left Australia) and returning to their respective homes. Pictures from the observing expedition are shared in Figure~\ref{figure:aus_people}.

In addition to successfully providing valuable engineering data on the equipment performance, and scientific data for pipeline development and subsequent analysis, the 2023~expedition also offered an opportunity for engagement, both internally and with the public. For many of the 10-person team, it was their first time witnessing totality, providing them with an unforgettable experience that would help mold how they talked to the other CATE~2024 participants and the general public about the project. Participants got field experience during an actual total eclipse and spent a significant amount of time interacting with crowds in Exmouth, strengthening their scientific communication skills. This helped foster excitement and engagement in the following months leading up to the 2024~eclipse. The team recorded their experiences with interviews and candid footage to produce a short video used both for engagement and for recruitment of State Coordinators and Lead Trainers as well as local team participants. This footage also contributed to a 26-minute documentary of the entire CATE~2024 project released in mid-2025 (see Section~\ref{subsec:afterengagement}). The team also live-streamed the eclipse experience over YouTube\footnote{\url{https://www.youtube.com/watch?v=-jdW4BUptGo}} to viewers across the world.

An additional field campaign for core team members occurred during the 2023~annular solar eclipse (ASE). Two teams observed the eclipse from Albuquerque, NM (DeForest, Kovac, and Davis; annular eclipse) and Loveland Pass, CO (Caspi, Seaton, Tosolini, and Zietlow; partial eclipse) using a 3D-printed coronagraph attachment on the telescopes. In addition to demonstrating the feasibility of using a coronagraph and high-altitude observations to detect the corona during non-total eclipses, the campaign provided another opportunity to beta-test procedures and equipment. Details of this expedition and coronagraph can be found in \citet{Seaton2024} and \citet{DeForest2024}.

\subsection{Coordinator Workshop}
All Regional and State Coordinators and Lead Trainers were required to attend a training workshop in January~2024, hosted at SwRI in San Antonio,~TX. This two-day workshop was modeled after the Spring~2023 workshop, and held over a weekend to reduce impact on school and work schedules for participants. Our major goal was to help the State Coordinators and Lead Trainers feel comfortable talking about CATE~2024, setting up the telescopes, and running their own training workshops for local teams. During the workshop, attendees specifically learned about total solar eclipses, the science of CATE, and the importance of community scientists.  They additionally received science communication, community engagement, and anti-harassment training, as well as ample time to practice setting up the telescopes and run through the eclipse-day procedure.

\subsection{Local Team Workshops}
\begin{figure}[H]
    \centering
    \centerline{\includegraphics[width=\textwidth]{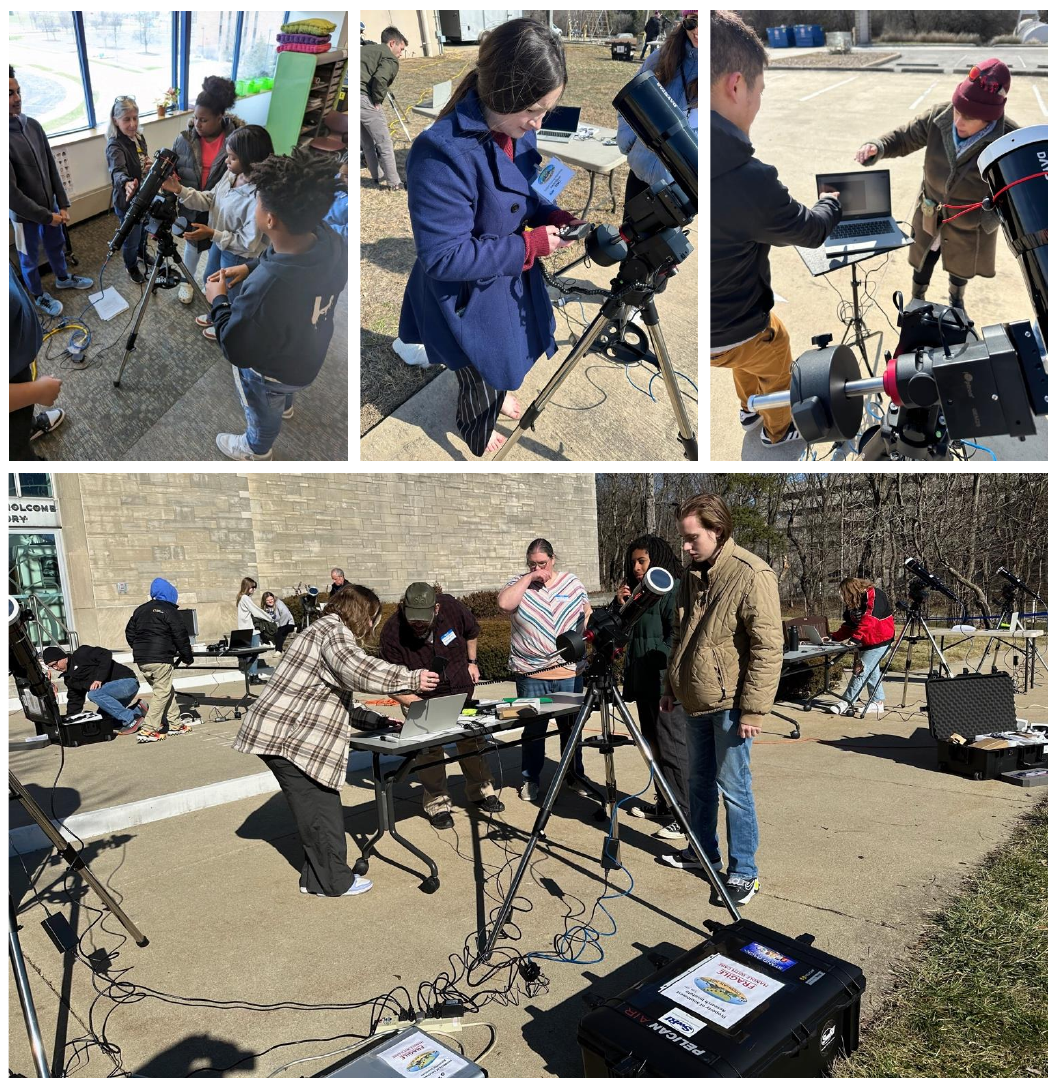}}
    \caption{Photographs from various CATE~2024 training workshops. (top left) SC Peggy Hill training middle school students at One City School; (top middle) LT Kayla Olson at the San Antonio Coordinator training workshop; (top right) San Antonio Coordinator training workshop; (bottom) IN training workshop.}
    \label{figure:workshop_photos}
\end{figure}

After attending the Coordinator workshop, State Coordinators returned home to host local team workshops. These workshops occurred in the early spring, typically at the State Coordinators' home institutions, or somewhere within relatively equal driving distance for all teams attending. These local workshops followed the same format as the Coordinator Workshop, typically held over a weekend for two days. All materials were shared with teams via Google Drive, and core CATE~2024 team members participated virtually for support. Team equipment was freight-shipped to each State Coordinator and distributed at these workshops. Participants practiced with the equipment, and took possession of it after the workshop. Some photos from these workshops are shown in Figure~\ref{figure:workshop_photos}. 

\subsection{Practice Sessions}
Three 'formal' practice sessions were slated for the weekends of 24~Feb, 9~Mar, and 23~Mar~2024. 6~Apr~2024 was left as a last-chance optional practice day for teams, with a mandatory practice on 7~Apr~2024, the day before the eclipse, preferably at their site location. Practice sessions were intentionally set on Saturdays to minimize the impact on school and work conflicts. The advantage of all teams practicing on the same day is that we could compare observations between teams, providing a valuable way to assess the quality of the observations. It was emphasized to teams that it was more important to practice repeatedly than to practice on these specific days. Core CATE~2024 team members were available on these days for technical support.

On these practice dates, we simulated eclipse observations by having volunteers observe from their planned eclipse day sites (as feasible) and operating with realistic procedure timing. Teams followed the entire observing procedure and collected all data as if the eclipse were happening under the real-time conditions of impending totality. Volunteers uploaded a specific subset of data (select GPS files, calibration files, focus images, and HDR sequences) to a team repository, allowing the Science Team to evaluate telescope polar alignment, camera angle alignment, object-centering (described in Section~\ref{sec:alignment}), and focus quality (described in Section~\ref{sec:focus}). We used these metrics to provide meaningful feedback to teams on success or required improvements following their practice sessions. Feedback was provided within a week of participants uploading their practice data to Google Drive.

Spring~2024 was quite cloudy along the path of totality. This made it difficult for teams to practice polar alignment, one of the more intricate steps of the procedure that requires clear skies with good visibility of the Sun. Roughly 75\% of teams were able to get at least three full on-sky practice sessions in where they received feedback from the core team. All teams were able to get in multiple practice sessions thanks to their dedication to observe whenever they had an opportune chance. Teams were instructed to set up indoors when there was poor weather and practice setting up, taking calibration data, mimicking totality, tear-down and packing of equipment, etc. However, while indoors, teams could not practice polar alignment. Multiple Regional and State Lead Trainers went out of their way to assist teams in-person outside of the scheduled practice dates. 

\subsection{Team Communication}
With such a large number of participants (over~250), effective communication between the core team, Coordinators and Trainers, and local team members was invaluable. CATE~2024 created a Slack workspace which allowed for conversation regarding many different aspects of the project, with the most active channel focused on questions/comments/concerns about the procedure. Throughout practice sessions, teams could ask questions that could be answered by other local team members, or by core team members. An advantage of building a tight-knit community was that local teams would answer other local teams' questions when they had the same problem. It made this channel a great resource as people could search keywords for a particular problem and it was very likely someone else also had that problem, allowing commonly asked questions to be quickly addressed. This helped build collaboration and camaraderie across all CATE~2024 teams, not just those that knew each other from the local workshops. 

\section{Eclipse Day: 8 April 2024}\label{sec:eclipse_day}
\subsection{Team Participation and Data Coverage}
CATE~2024 teams were provided with a phone tree to call other teams if they were having any equipment or software concerns. This was broken down by region and time, such that teams who needed assistance would not be calling teams who were in the middle of totality. This was another benefit of the stair-stepped responsibilities for Coordinators and Trainers, such that no one person was inundated with questions. Weather conditions along the path of totality were mostly clear for a majority of CATE~2024 sites at the local time of totality. Unfortunately, adverse weather did impact data collection for teams in southern Texas and New York. Despite these conditions, 100\% of CATE~2024 teams participated on eclipse day, making observations near-continuous, with short gaps in southern Texas and New York. 

\subsection{Public Engagement}\label{subsec:engagement}
To extend CATE~2024 education and engagement on the day of the eclipse, NSF NCAR Education, Engagement \& Early-Career Development (EdEC), in collaboration with SwRI, CU/LASP, and the UCAR Center for Science Education, hosted activity tables at three local community festivals along the path of totality. The team collaborated with local hosts and community festivals, many of which had been planned well in advance of the eclipse, to tap into local knowledge of community interests and practical considerations for organizing large-scale events. This allowed the CATE~2024 project team to have more authentic and meaningful interaction with the community and to host tables at these already-planned events.

The three locations chosen were: 1) SoMa in the Dark: Path of Totality Eclipse Watch Party in Little Rock, Arkansas (O3-AR-041); 2) Total Eclipse of the Park in Sikeston, Missouri (O4-MO-042); and 3) Holcomb Observatory and Planetarium at Butler University in Indianapolis, Indiana (O5-IN-043). These locations were chosen because the festivals were free to the public, had hosts who were excited to have us, and served a range of local communities. A lack of presence from other science organizations with engagement activities was also a factor, so that we could help fill gaps along the path of totality. Based on numbers provided by the organizers on the total number of attendees, we estimate engaging about 4,000~local community members at our activity tables across all three locations. Additionally, staffing these activity tables served as an excellent professional development opportunity for early-career scientists interested in practicing their science communication skills with general public audiences. Thus, we paired postdocs with each of the community engagement specialists that staffed the tables.

At the activity tables, community members could view the solar telescope setups being used by the community scientist teams on the CATE~2024 project. Staff focused the telescopes on the Sun where they could discuss sunspots and eclipses with visitors at the table. We also emphasized that most team members involved in CATE~2024 had never used telescopes like these before and were now quite proficient with them, so there are many ways to be involved in science even if it feels daunting.

Visitors could also participate in two hands-on activities about why our Sun’s corona is of interest to scientists, polarization of light (one of the research objectives that made the CATE~2024 project unique compared to other community science projects), and ultraviolet radiation. These activities were designed with best-practices in mind \citep[e.g.,][]{Idema2019} where the engagement is focused on a single topic, interactive, playful, and for the whole family. In the first activity, participants created coronal chalk art and, in the second, they created and then experimented with UV-sensitive bead bracelets. Additionally, we created a bilingual postcard (English and Spanish) highlighting how scientists use total solar eclipses to study our Sun, the physical effects that one may experience during an eclipse, and how different cultures may have different ways in engaging with eclipses. Visitors could take any of the activities home. We also distributed certified ISO~12312-2 solar viewing glasses and pinhole viewers provided by both the NASA \textit{PUNCH} mission and NSF NCAR. Informally, we received feedback from festival hosts and community members that they were excited about having us at their festivals, highly enjoyed the activities, and were appreciative of being able to directly interact with community scientists and leading scientific organizations.

\subsection{Media Engagement}\label{subsec:media}
In addition to engaging with the public in-person, multiple images and videos from various CATE~2024 observation sites and engagement teams were highlighted across multiple media outlets around eclipse day. This allowed the greater public, both on and off the path, to see what the project team members were doing on eclipse day in real time. 

Using Meltwater analytics, which provides tools and services that help businesses monitor and analyze media coverage, social media conversations, and other online content, we were able to track the engagement of media products. Between 1~March~2024 and 14~April~2024, there were 121~media products -- content or output created by media organizations, such as articles, videos, broadcasts, or social media posts -- that included CATE~2024. Each media product had varying levels of ``reach,'' referring to the estimated number of unique individuals who have been exposed to a particular piece of media content, such as an article or social media post. It is a measure of the potential audience size that could have seen or interacted with the content. The project used reach as a metric to gauge the impact or influence of media coverage, indicating how widely the content has been distributed and viewed across different platforms. The top performing article, ``Americans shared a powerful moment during 2024~solar eclipse'' \citep{usa_today2024}, had an estimated reach of over 200~million. 

\section{Post-Eclipse Engagement Activities}\label{subsec:afterengagement}
In a collaborative effort, NSF NCAR EdEC and a scientist from UCAR COSMIC worked to develop a series of educational activities centered around CATE~2024 telescope use. These activities are designed to be adaptable across elementary, middle, high school, and college levels, aligning with Next Generation Science Standards \citep{ngss2013} when applicable. 

These activities focus on key astronomical concepts such as sunspot observation, planetary exploration, lunar comparisons between Titan and our Moon, and celestial navigation using stars. As the grade level changes, activities transition from structured, cookbook-type laboratory experiences to more inquiry-based learning with in-depth pre- and post-lab questions and extension exercises. Importantly, these activities can be implemented not only in traditional classroom settings, but also in community and informal learning environments. By providing a flexible framework and adaptable materials, the project aims to extend the utility of the CATE~2024 telescopes beyond the initial CATE~2024 experience and support ongoing astronomical education and research.

Throughout the duration of CATE~2024, we also produced a PBS-style documentary. The film highlights not only the CATE~2024 project, but also community member perspectives on total solar eclipses and being involved in a community science campaign. In addition to the filmmaker on the core CATE~2024 team (Zietlow), we collaborated with four filmmakers in three different communities along the path of totality to co-create short segments for inclusion into the documentary so that we could feature a wide variety of experiences from across the geographic spread of the project. The documentary, \textit{Gathered in Darkness}, is available for screenings on request\footnote{\url{https://edec.ucar.edu/public/field-projects/cate-2024}} and is a 2025 Public Media Awards honoree. Another video project about CATE~2024 was also created in collaboration with AGU~TV\footnote{\url{https://youtu.be/EAH9oqrfQc8}}.

\section{Observations and Data Processing}\label{sec:results}
Once data was received from teams (both via Google Drive and physically mailing thumb drives), preliminary data processing started. To maximize the number of images that could be captured and saved during totality, data was initially saved as TIFF images with metadata in separate files. The first processing step is to reformat the TIFF files and all needed metadata into FITS files with proper header information, following FITS~4.0 standards. These FITS files are then used throughout the rest of the data pipeline. Preliminary data products from a handful of CATE~2024 sites are shown in this section, with scientific analysis of the full dataset left for a future publication.

\subsection{High Dynamic Range Images}
For each of the eight exposure times (see Section~\ref{sec:pg_science}) in the totality sequence, a series of dark images were taken with the same exposure time. For each site, a total of 160~dark frames were collected, resulting in 20~dark images for each exposure time. A median of the 20~dark images is used to create a master dark for each exposure time. The master dark image is then subtracted from the corresponding totality sequence images with the same exposure time for that site.

\begin{figure}[H]
    \centering
    \centerline{\includegraphics[width=\textwidth]{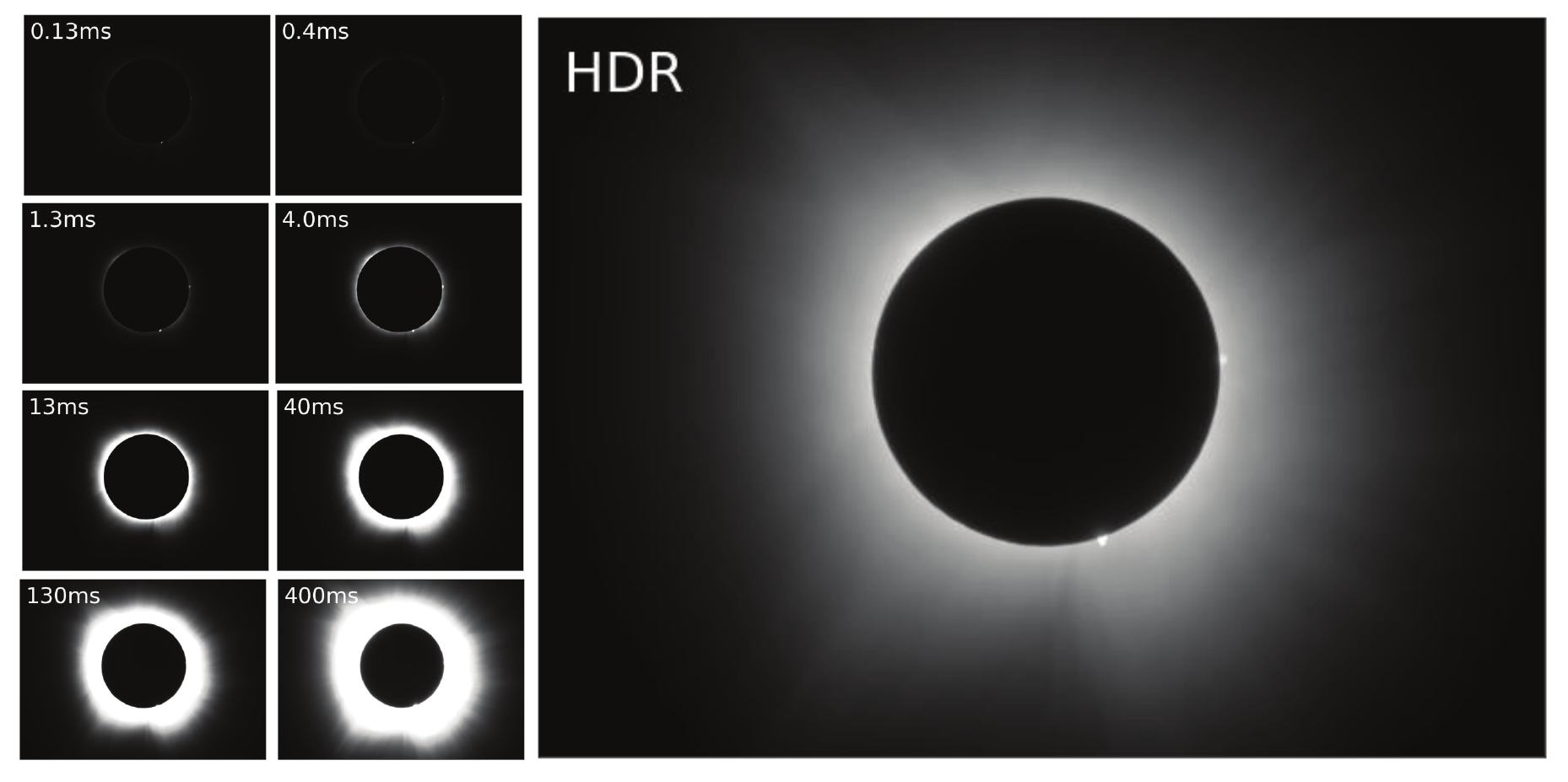}}
    \caption{Individual exposures and compiled HDR image from Site~016 (UTC~19:00:35). Individual exposures are shown in the smaller images, with the exposure time in the top left corner. The large image on the right is the result of combining those exposures into an HDR.}
    \label{figure:hdr}
\end{figure}

Prior to distributing cameras, the science team collected in-house calibration data to fully characterize the performance of each camera, including polarimetric accuracy and determining its optimal linear range, to be discussed in detail in a future science paper (including incorporation of flat-field images). HDRs are composed of sets of dark-subtracted images at multiple exposures. For a given exposure, only the pixels that fall within the linear regime of the camera are included in the final HDR. This means that for each pixel, only a subset of the eight exposures are used to create the HDR, with a minimum of one exposure time per pixel. Dark-subtracted pixels that contribute to the HDR are first normalized by their exposure time. These pixels are then co-added and normalized by the number of exposures that contribute to each respective HDR pixel. An example exposure sequence and HDR composite image are shown in Figure~\ref{figure:hdr}. 

\subsection{Polarized Brightness and Image Co-alignment}
\label{subsec:pb}
\begin{figure}[H]
    \centering
    \centerline{\includegraphics[width=\textwidth]{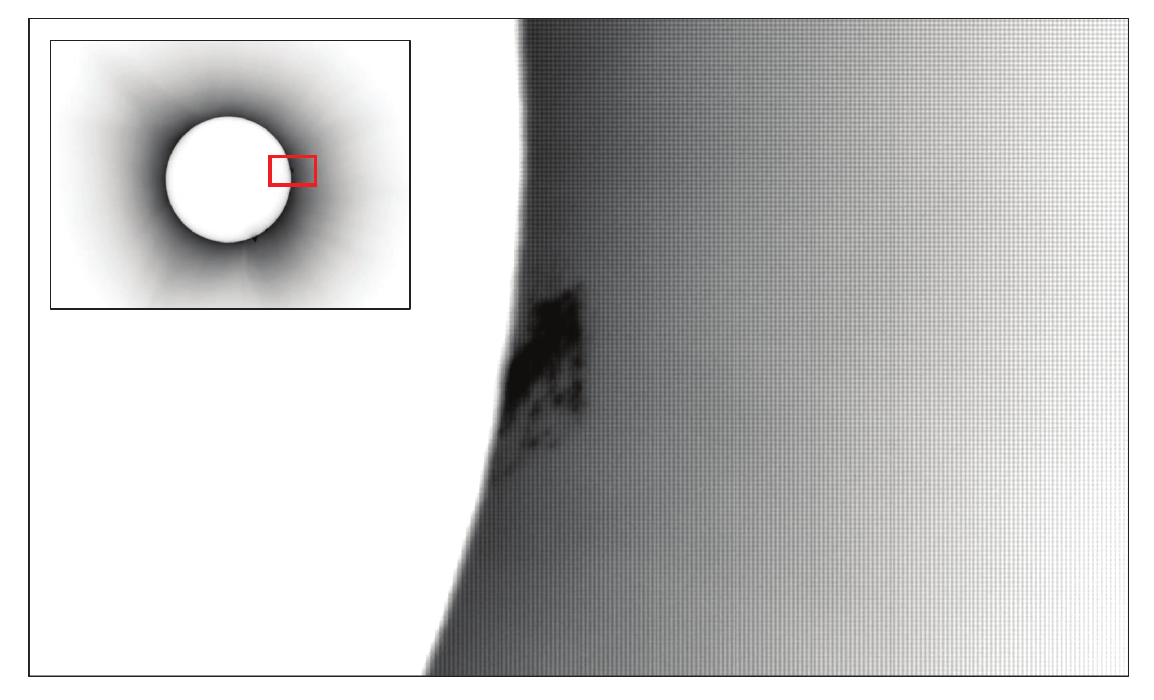}}
    \caption{HDR image from Site~034 (UTC~19:33:23) displayed with an inverse color table. The inset image is the full field of view, while the large image emphasizes ``crosshatch'' of the individual polarizers within the macropixels of the region of interest outlined in red.}
    \label{figure:hdr_roi}
\end{figure}

\begin{figure}[H]
   \centering
   \centerline{\includegraphics[width=\textwidth]{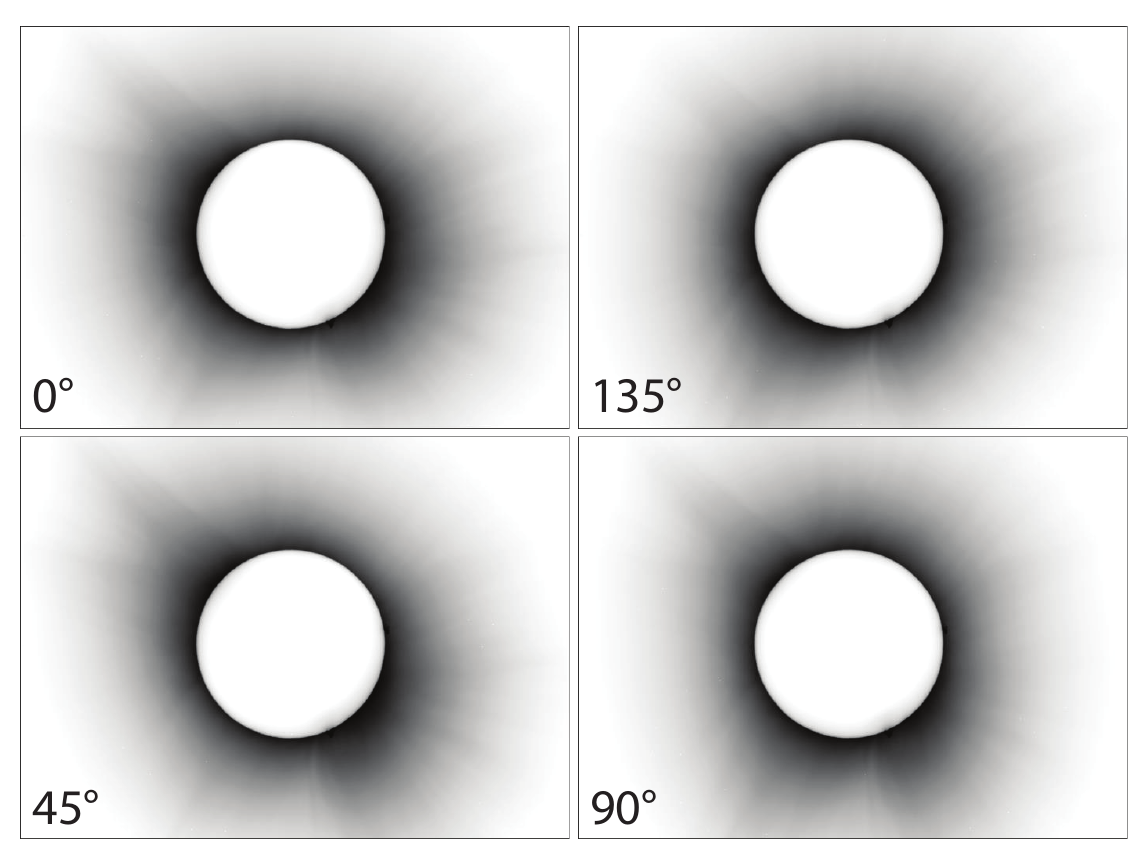}}
  \caption{Four individual polarization states ($0\degree$, $45\degree$, $90\degree$, $135\degree$) from Site~034, extracted from the master image displayed in Figure~\ref{figure:hdr_roi}, displayed with an inverse color table.}
  \label{figure:hdr_pol}
\end{figure}

The computed HDRs are a composite of all four polarization states. Figure~\ref{figure:hdr_roi} highlights an inset of the HDR to emphasize the ``crosshatch'' appearance which is a result of the four polarization state subpixels creating a macropixel. To calculate polarized brightness, we separated each HDR image into its individual polarization channels, shown in Figure~\ref{figure:hdr_pol}, each offset from the others by $45\degree$. For simplicity, during this preliminary analysis, we calculate the quantity $^\circ pB$, which is the total polarized light independent of direction. $^\circ pB$ can be computed directly from a four-angle observation such as ours using the formula in the Appendix of \citet{DeForest2022}, $^\circ pB = (Q^2 + U^2)^{1/2}$, where $Q$ and $U$ are the Stokes parameters in the instrument frame, given by $Q = B_{|} - B_{-}$ and $U = B_{\backslash} - B_{/}$ where $B_\theta$ represents the radiance through a polarizer at a given angle, corresponding to each one of the four polarization measurements ($0\degree$, $45\degree$, $90\degree$, $135\degree$) made by the CATE~2024 camera. Note that in the low corona the distinction between $^\circ pB$ and the true tangential polarization $^\perp pB$ is negligible because the polarization is almost entirely tangential to the limb anyway.

To generate co-aligned movies of individual $^\circ pB$-resolved CATE~2024 data, it is necessary to remove any residual drift due to inexact telescope alignment and any residual rotational misalignment as well. We achieved this by first identifying the location of the Moon using the limb-fitting algorithm \textsf{p2sw\_fit\_limb.pro}, which is released in \textsf{SolarSoft IDL} as part of the \textit{PROBA2} SWAP data calibration package \citep{Seaton2013SWAP}. For the preliminary examples shown here, we determined the rotation angle of each individual set of observations from a given site by comparison to an external observation with known orientation from the \textit{Solar Dynamics Observatory} (SDO) Atmospheric Imaging Assembly \citep{Lemen2012}.

\begin{figure}
   \centering
   \centerline{\includegraphics[width=\textwidth]{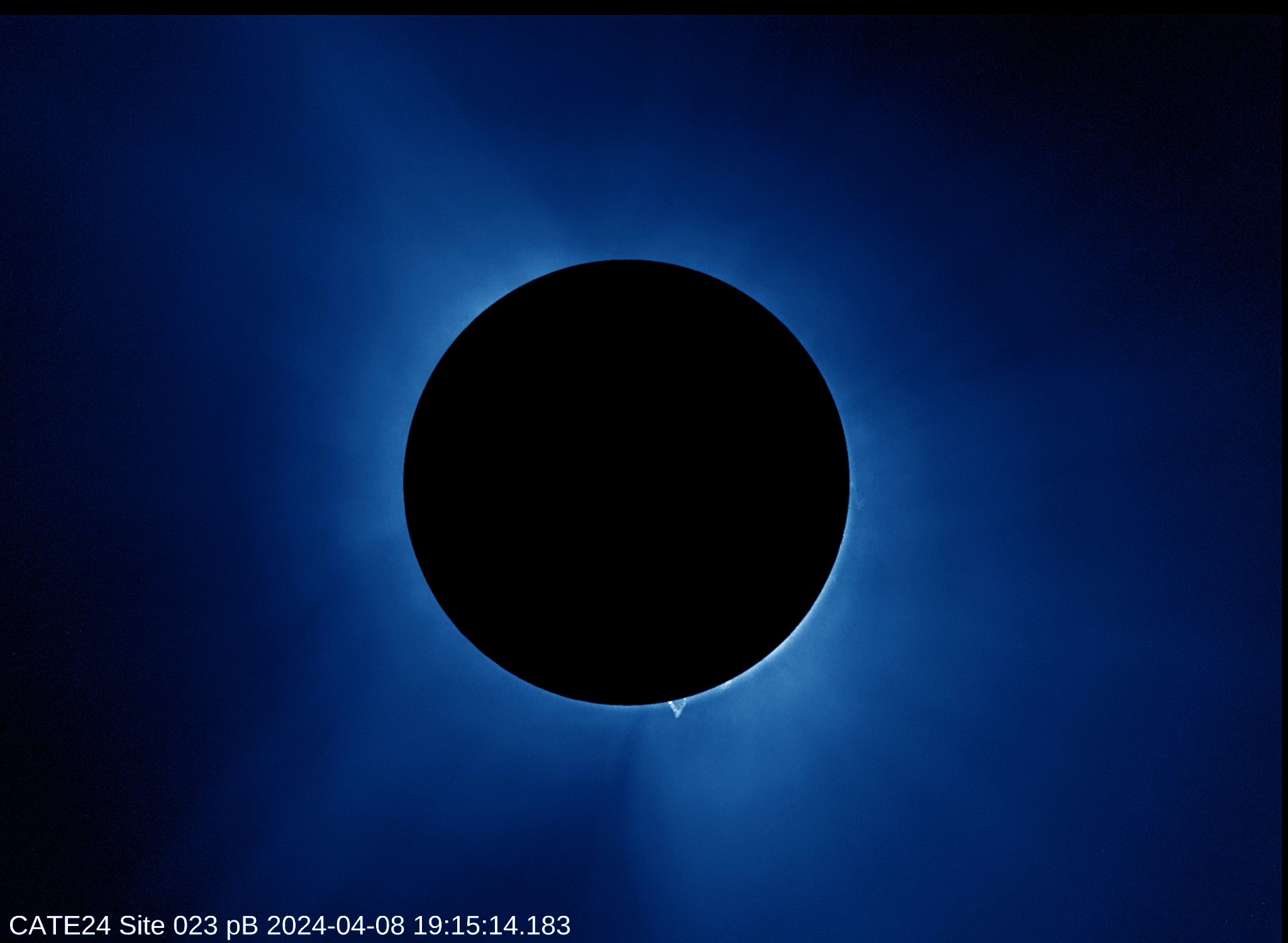}}
  \caption{Polarized brightness ($^\circ pB$) image from Site~023. See supplementary material for a movie including Sites 008, 016, 023, and 034.}
  \label{figure:pb_023}
\end{figure}

For the corona to be aligned across multiple images, we must co-align to the unseen center of the Sun, hidden by the Moon. Knowing the position of the Moon, rotation angle, image pixel scale, and the location and time of the observation is sufficient to fully define the World Coordinate System (WCS) metadata structure to define the image projection in heliographic coordinates \citep[see][for details]{Thompson2006}. Once defined, we can use the WCS information to compute the location of Sun-center in each processed $^\circ pB$ image, from which we can re-interpolate the images to a consistent, Sun-centered frame. Figure~\ref{figure:pb_023} shows a processed $^\circ pB$ image from Site~023 included in our preliminary analysis, and the accompanying animation shows the co-aligned preliminary movie from Sites~008, 016, 023, and 034. 

\subsection{RGB Polarization Processing}
\label{subsec:rgb}
\citet{DeForest2022} provide a framework for converting three-angle polarimetric measurements to red-green-blue (RGB) color space where hue visually reflects polarization direction and color saturation reflects the polarization fraction \citep[see][for details]{patel2023}. While our $^\circ pB$ images reveal the total polarized brightness without respect to angle, these processed images provide a visualization from which it is possible to understand the structure of the corona and the complete characteristics of its linear polarization in a single image. 

HDR images separated into the four polarizer angles shown in Figure~\ref{figure:hdr_pol} are first converted to a 3-polarizer basis with polarizer angles separated by 60$^\circ$ -- i.e., --60$^\circ$, 0$^\circ$ and +60$^\circ$ -- using \texttt{solpolpy}, a solar polarization resolver \citep{hughes_2024, Patel24}. To balance the large-scale intensity gradients in the radial direction, a radial filter algorithm is applied, \texttt{p2sw\_image\_filter.pro}, developed for \textit{PROBA2}/SWAP \citep{Seaton2023}. This makes the coronal structures more visually uniform. To enhance the visibility and contrast of fine-scale coronal structures, we then applied the Multi-scale Gaussian Normalization (MGN) technique \citep{Morgan2014SoPh} to these three individual images. Outlier values and NaNs are filtered before normalizing the intensity of the three images. The three enhanced images are added to the intensity-scaled images at 0$^\circ$ and $\pm$60$^\circ$, yielding single images that reveal both the large-scale structure of the corona and fine-scale features in a single, well-normalized frame.

\begin{figure} 
\centering
\centerline{\includegraphics[width=\textwidth]{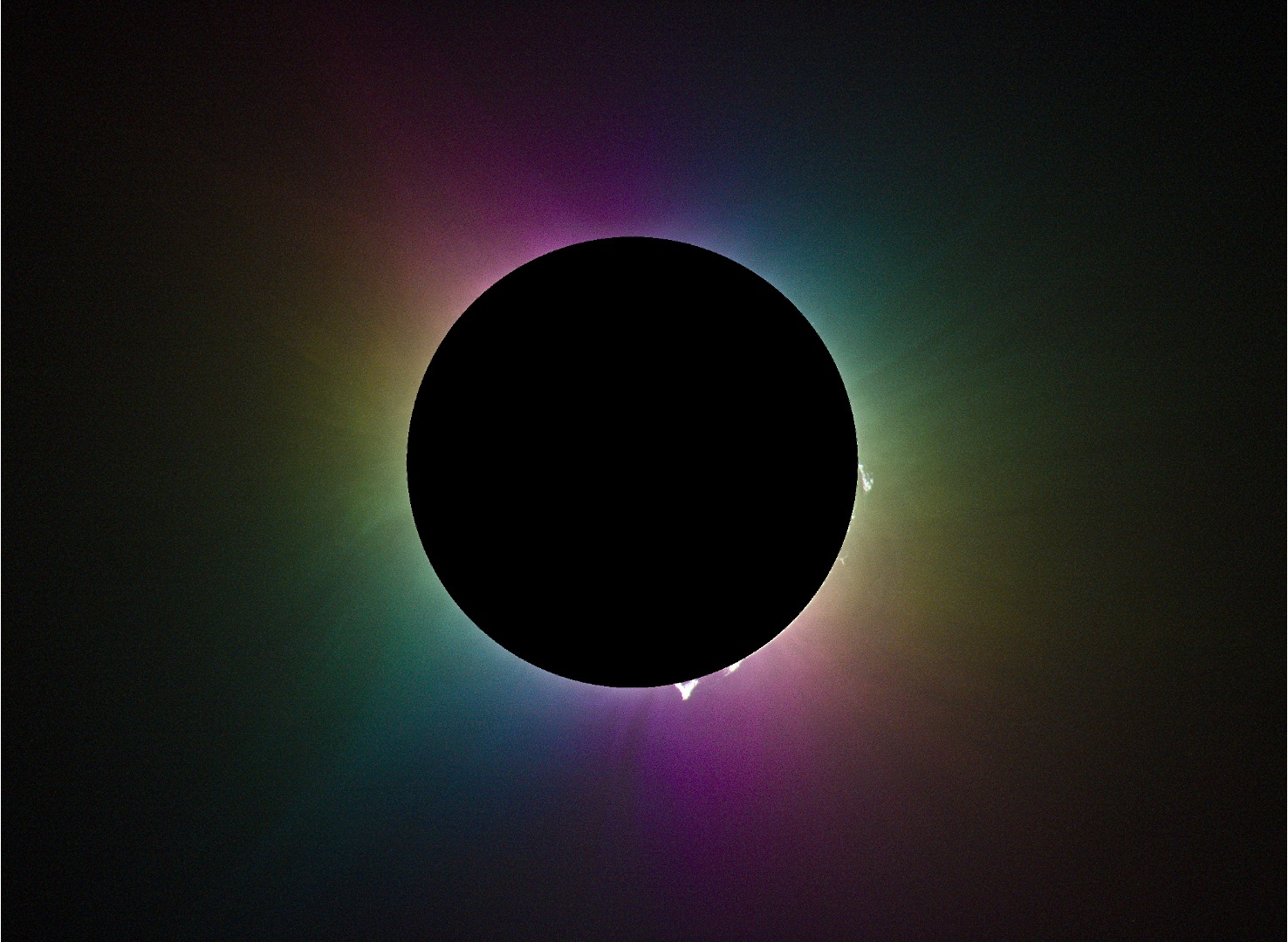}}
    \caption{Linearly polarized broadband visible-light K-corona from one HDR from Site~023, from the same underlying data as Figure~\ref{figure:pb_023}, with polarization angle encoded as color and polarization fraction as color saturation using the method of \citet{patel2023}.}
    \label{figure:rainbow}
\end{figure}

Just as different colors can be represented as a linear combination of $r$, $g$, $b$ primary colors, different polarizations can be represented as a combination of three polarization measurements at three polarizer angles. So, the multi-scale-processed composite images are then passed into the red, green and blue channels of a single RGB image to produce a color image showing fraction and direction of polarization in the corona (Figure~\ref{figure:rainbow}). In such RGB polarization images, the color saturation represents the polarization fraction (i.e., whiter is less overall polarized) and hue represents the polarization angle. The colors progress through the spectrum from red to violet, with red corresponding to a 0$^\circ$ polarization angle. Moving counter-clockwise around the image, the color progression represents the change in polarization angle as a function of solar azimuth. The color pattern repeats after 180$\degree$, since polarization has an 180$\degree$ ambiguity.

We note that prominence structures visible at the limb of the Sun appear white. This does not mean they are saturated in original images (e.g., Figure~\ref{figure:pb_023}), but rather these structures are observed primarily in emission, not Thomson scattering, and thus are largely unpolarized, hence appearing white. The inner corona is less polarized than the outer corona; this is because the polarization depends on scattering angle, and the large solid angle the photosphere subtends at low heights allows scattering over a large range of angles \citep[see][both Chapter 6, Section B and Appendix IV]{Billings_1966}. Correspondingly, there is less color saturation in the inner corona than at larger heights. The image in Figure~\ref{figure:rainbow}was taken just before third contact, so the southwest region of the image shows lower parts of the corona than the northwest region and thus the lower right is less saturated upper left, making opposite sides of the lunar disk vary in saturation. 

\section{Discussion and Conclusions}\label{sec:discussion}  
The 8~April 2024~TSE that traversed North America made totality potentially visible to over 44~million people who lived along the path, as well as many others who traveled to the shadow \citep{space2024eclipse}. The Citizen CATE~2024 project engaged with local communities along the path of totality in the United States, giving community participants the opportunity to contribute to a unique astronomical dataset. With over 250~total participants across 43~teams, CATE~2024 reached into many underserved communities both on and off the path. 

The original 2017 Citizen CATE Experiment \citep{Penn2020} showed that a large-scale participatory science project was achievable, and that recruitment and training of individuals with no prior astronomy experience could result in a novel dataset with meaningful applications. Following in their footsteps, the CATE~2024 team adapted this framework and evolved new project goals for the next generation of upcoming scientists, including new equipment and using a camera sensitive to polarized light. Teams were composed of people from various different backgrounds, from students through professional astronomers, with strong representation from underserved groups such as women and minorities, as well as participation from Indigenous communities. Ten students, both undergraduate and graduate, filled leadership roles on a large-scale science project, providing exposure to many aspects of STEM careers. 

100\% of teams participated on eclipse day, and over 85\% of the 36~community-led CATE~2024 observing teams successfully collected data, for a combined $\sim$47,000~images of totality, amounting to over 50~minutes of polarized images of the inner and middle solar corona. Data gaps along the path ($\sim$15\%) were due to poor weather conditions, not personnel or equipment failure. Future analysis of these observations will contribute to a better understanding of many aspects of the corona and solar wind, including characteristics of connectivity and reconnection in the middle corona and nascent solar wind flows. 

The rich CATE~2024 dataset also allows for unique analyses in combination with data from other observational datasets (such as those at different wavelengths, \citep[e.g.,][]{rivera2024} and physics-based magnetohydrodynamic (MHD) modeling \citep[e.g.,][]{downs2025}. CATE~2024 data can be compared to other ground-based data collected in 2024 from other participatory science projects like DEB, Eclipse Mega\-Movie, and Sun\-Sketcher, as well as space-based observations from missions such as GOES, Solar Orbiter, SDO, and Parker Solar Probe. Spacecraft data, in particular, can provide views of the corona in wavelengths not observable from the ground, like EUV, and instruments like SDO/HMI provide information on the magnetic field via magnetograms. Airborne observations with the NASA WB-57 provide data in the MWIR \citep{seaton2024_agu}, which will help address unanswered questions about the nature of coronal emission at these wavelengths. 

The CATE~2024 dataset is currently being formatted in compliance with FITS~4.0 standards, compatible with \textsf{Astropy} and \textsf{SunPy}. The full, raw dataset is $\sim$2~TB including calibration data. The data pipeline development and the full-length calibrated CATE~2024 movie are still in progress. The fully calibrated dataset and accompanying data reduction routines will be available to the public and linked via GitHub in a future science paper. 

The CATE~2024 Coordinators continue to meet 1--2 times per month to provide updates, to discuss further education and public outreach activities and how they have been using the telescopes post-eclipse. Additional outreach materials are being developed for both daytime and nighttime observing. With the equipment now housed in community-accessible locations like schools and libraries, many groups have continued to use the equipment for various outreach and public engagement activities, including the March 2025~lunar eclipse. 

Totality will not again be visible from the continental United States until~2044. There are over a dozen more TSEs that occur in various parts of the world between now and then, including several over large swaths of land. Future distributed eclipse observations, both from the air and from the ground, will be essential to continue expanding our understanding of the solar corona. Projects like CATE~2024 demonstrate the feasibility and high value of large-scale participatory science projects. These projects empower people with and without formal scientific training to actively contribute to novel scientific research, further expanding the reach and scientific efforts by increasing data collection, increasing public scientific literacy, broadening access to science (particularly in underserved communities), and fostering collaboration among folks who may otherwise never cross paths. All of these reasons, and more, contribute to the growing impact of participatory science projects, which can be applied not just in astronomy, but across many disciplines. 

\begin{acks}
Funding for Citizen CATE~2024 was provided by grants from the National Science Foundation (award numbers 2231658, 2308305, 2308306, 2511904, 2511905, \& 2511907), NASA (grant numbers 80NSSC21K0798 \& 80NSSC23K0946), and SwRI (internal award R6395). We thank Teledyne FLIR LLC and Daystar Filters LLC for their generous contributions and assistance in procuring critical equipment affordably and within tight schedule constraints. We thank Brunton and StellarVue for equipment discounts, and the SwRI Purchasing and Travel departments and Astro Trails Ltd for negotiating additional discounts and assisting with complex logistics. This material is based upon work supported by the NSF National Center for Atmospheric Research, which is a major facility sponsored by the U.S. National Science Foundation under Cooperative Agreement No.~1852977. We thank Matthew Penn and Robert Baer for pioneering the original 2017 Citizen CATE Experiment, which invaluably contributed to the success of CATE~2024.
\end{acks}

\begin{authorcontribution}
S.A.~Kovac led the writing of the manuscript and led CATE~2024 project management. 

A.~Caspi and D.B.~Seaton contributed substantially to the manuscript, and led the overall CATE~2024 mission and science, respectively. 

A.~Caspi, S.A.~Kovac, D.~Lamb, R.~Patel, D.B.~Seaton, and  A.~Tosolini contributed to data processing, calibration, and data products. P.~Bryans, S.~Gosain, J.~Jackiewicz, V.~Klein, and K.~Reardon made additional data contributions.

A.~Ursache led development of the SolarEclipseApp custom observing software, and contributed to calibrations and debugging. 

W.~Reed led CATE~2024 communication efforts and D.W.~Zietlow led public engagement efforts; both contributed to various communication, engagement, and media sections of the manuscript. J.R.~Burkepile, C.E.~DeForest, R.~Haacker, and J.K.~Williams made additional contributions to engagement efforts.

S.J.~Davis and S.A.~Kovac led development and implementation of the observing procedure and participant training.

D.~Elmore, M.~Hill, S.A.~Kovac, A.~Ursache, and P.~Yanamandra-Fisher contributed experience and lessons learned from involvement with the original Citizen CATE 2017 Experiment and other prior eclipse observations, including with polarization-sensitive cameras.

The CATE~2024 Coordinators and Trainers listed in Table~\ref{table:roles} led coordination, training, and support for community participants.

All authors contributed to logistics and/or data collection, and to manuscript review.
\end{authorcontribution}

\begin{fundinginformation}
Funding for Citizen CATE~2024 was provided by grants from the National Science Foundation (award numbers 2231658, 2308305, 2308306, 2511904, 2511905, \& 2511907), and NASA (grant numbers 80NSSC21K0798 \& 80NSSC23K0946). S.A.~Kovac and A.~Caspi were also partially supported by SwRI (internal award R6395). R.~Patel is supported by NASA H-USPI grant 80NSSC21K1860 to carry out polarimetric analysis. Summer student contributions through the Boulder Solar Alliance Research Experience for Undergraduates Program were funded by the National Science Foundation (CATE~2024 awards, and award 1950911).
\end{fundinginformation}

\begin{dataavailability}
CATE~2024 raw and calibrated data will be available via public-facing web servers at SwRI and partner institutions, and will be archived for long-term accessibility in appropriate federally maintained data repositories such as data.gov. We acknowledge the data sovereignty of the communities with which we worked during this project and implement equitable data sharing guided by the CARE Principles for Indigenous Data Governance.
\end{dataavailability}

\begin{materialsavailability}
Nearly all Citizen CATE~2024 equipment is available off the shelf, with the exception of the 3D-printed diffuser holder, plans for which can be found at {\url{https://www.tinkercad.com/things/bkvEqK36zIt-telescope-bracket-v2-all}}.
\end{materialsavailability}

\begin{codeavailability}
Data reduction routines will be available to the public and linked via GitHub in a future science paper. 
\end{codeavailability}

\begin{ethics}
\begin{conflict}
The authors declare that they have no conflicts of interest or competing interests.
\end{conflict}
\end{ethics}

\bibliographystyle{spr-mp-sola}
\bibliography{references}  

\end{document}